\documentclass[11pt]{article}
\usepackage{lettrine}
    \usepackage[breakable]{tcolorbox}
    \usepackage{parskip} 
    \usepackage{iftex}
    \ifPDFTeX
    	\usepackage[T1]{fontenc}
    	\usepackage{mathpazo}
    \else
    	\usepackage{fontspec}
    \fi
\usepackage{fancyhdr}
    \usepackage{graphicx}
    
    \usepackage{caption}

    \usepackage{float}
    \usepackage{xcolor} 
    \usepackage{enumerate} 
    \usepackage{geometry} 
    \usepackage{amsmath} 
    \usepackage{amssymb} 
    \usepackage{textcomp} 
    \AtBeginDocument{%
    }
    \usepackage{upquote} 
    \usepackage{eurosym} 
    \usepackage[mathletters]{ucs} 
    \usepackage{fancyvrb} 
    \usepackage{grffile} 
    \makeatletter 
    \@ifpackagelater{grffile}{2019/11/01}
    {
    }
    {
      \def\Gread@@xetex#1{%
        \IfFileExists{"\Gin@base".bb}%
        {\Gread@eps{\Gin@base.bb}}%
        {\Gread@@xetex@aux#1}%
      }
    }
    \makeatother
    \usepackage[Export]{adjustbox} 
    \adjustboxset{max size={0.9\linewidth}{0.9\paperheight}}

    \usepackage{hyperref}
    \usepackage{titling}
    \usepackage{longtable} 
    \usepackage{booktabs}  
    \usepackage[inline]{enumitem} 
    \usepackage[normalem]{ulem} 
    \usepackage{mathrsfs}

    \definecolor{urlcolor}{rgb}{0,.145,.698}
    \definecolor{linkcolor}{rgb}{.71,0.21,0.01}
    \definecolor{citecolor}{rgb}{.12,.54,.11}

    \definecolor{ansi-black}{HTML}{3E424D}
    \definecolor{ansi-black-intense}{HTML}{282C36}
    \definecolor{ansi-red}{HTML}{E75C58}
    \definecolor{ansi-red-intense}{HTML}{B22B31}
    \definecolor{ansi-green}{HTML}{00A250}
    \definecolor{ansi-green-intense}{HTML}{007427}
    \definecolor{ansi-yellow}{HTML}{DDB62B}
    \definecolor{ansi-yellow-intense}{HTML}{B27D12}
    \definecolor{ansi-blue}{HTML}{208FFB}
    \definecolor{ansi-blue-intense}{HTML}{0065CA}
    \definecolor{ansi-magenta}{HTML}{D160C4}
    \definecolor{ansi-magenta-intense}{HTML}{A03196}
    \definecolor{ansi-cyan}{HTML}{60C6C8}
    \definecolor{ansi-cyan-intense}{HTML}{258F8F}
    \definecolor{ansi-white}{HTML}{C5C1B4}
    \definecolor{ansi-white-intense}{HTML}{A1A6B2}
    \definecolor{ansi-default-inverse-fg}{HTML}{FFFFFF}
    \definecolor{ansi-default-inverse-bg}{HTML}{000000}

    \definecolor{outerrorbackground}{HTML}{FFDFDF}

    \DefineVerbatimEnvironment{Highlighting}{Verbatim}{commandchars=\\\{\}}

    \let\Oldtex\TeX
    \let\Oldlatex\LaTeX
    \renewcommand{\TeX}{\textrm{\Oldtex}}
    \renewcommand{\LaTeX}{\textrm{\Oldlatex}}

\makeatletter
\def\PY@reset{\let\PY@it=\relax \let\PY@bf=\relax%
    \let\PY@ul=\relax \let\PY@tc=\relax%
    \let\PY@bc=\relax \let\PY@ff=\relax}
\def\PY@tok#1{\csname PY@tok@#1\endcsname}
\def\PY@toks#1+{\ifx\relax#1\empty\else%
    \PY@tok{#1}\expandafter\PY@toks\fi}
\def\PY@do#1{\PY@bc{\PY@tc{\PY@ul{%
    \PY@it{\PY@bf{\PY@ff{#1}}}}}}}
\def\PY#1#2{\PY@reset\PY@toks#1+\relax+\PY@do{#2}}

\@namedef{PY@tok@w}{\def\PY@tc##1{\textcolor[rgb]{0.73,0.73,0.73}{##1}}}
\@namedef{PY@tok@c}{\let\PY@it=\textit\def\PY@tc##1{\textcolor[rgb]{0.25,0.50,0.50}{##1}}}
\@namedef{PY@tok@cp}{\def\PY@tc##1{\textcolor[rgb]{0.74,0.48,0.00}{##1}}}
\@namedef{PY@tok@k}{\let\PY@bf=\textbf\def\PY@tc##1{\textcolor[rgb]{0.00,0.50,0.00}{##1}}}
\@namedef{PY@tok@kp}{\def\PY@tc##1{\textcolor[rgb]{0.00,0.50,0.00}{##1}}}
\@namedef{PY@tok@kt}{\def\PY@tc##1{\textcolor[rgb]{0.69,0.00,0.25}{##1}}}
\@namedef{PY@tok@o}{\def\PY@tc##1{\textcolor[rgb]{0.40,0.40,0.40}{##1}}}
\@namedef{PY@tok@ow}{\let\PY@bf=\textbf\def\PY@tc##1{\textcolor[rgb]{0.67,0.13,1.00}{##1}}}
\@namedef{PY@tok@nb}{\def\PY@tc##1{\textcolor[rgb]{0.00,0.50,0.00}{##1}}}
\@namedef{PY@tok@nf}{\def\PY@tc##1{\textcolor[rgb]{0.00,0.00,1.00}{##1}}}
\@namedef{PY@tok@nc}{\let\PY@bf=\textbf\def\PY@tc##1{\textcolor[rgb]{0.00,0.00,1.00}{##1}}}
\@namedef{PY@tok@nn}{\let\PY@bf=\textbf\def\PY@tc##1{\textcolor[rgb]{0.00,0.00,1.00}{##1}}}
\@namedef{PY@tok@ne}{\let\PY@bf=\textbf\def\PY@tc##1{\textcolor[rgb]{0.82,0.25,0.23}{##1}}}
\@namedef{PY@tok@nv}{\def\PY@tc##1{\textcolor[rgb]{0.10,0.09,0.49}{##1}}}
\@namedef{PY@tok@no}{\def\PY@tc##1{\textcolor[rgb]{0.53,0.00,0.00}{##1}}}
\@namedef{PY@tok@nl}{\def\PY@tc##1{\textcolor[rgb]{0.63,0.63,0.00}{##1}}}
\@namedef{PY@tok@ni}{\let\PY@bf=\textbf\def\PY@tc##1{\textcolor[rgb]{0.60,0.60,0.60}{##1}}}
\@namedef{PY@tok@na}{\def\PY@tc##1{\textcolor[rgb]{0.49,0.56,0.16}{##1}}}
\@namedef{PY@tok@nt}{\let\PY@bf=\textbf\def\PY@tc##1{\textcolor[rgb]{0.00,0.50,0.00}{##1}}}
\@namedef{PY@tok@nd}{\def\PY@tc##1{\textcolor[rgb]{0.67,0.13,1.00}{##1}}}
\@namedef{PY@tok@s}{\def\PY@tc##1{\textcolor[rgb]{0.73,0.13,0.13}{##1}}}
\@namedef{PY@tok@sd}{\let\PY@it=\textit\def\PY@tc##1{\textcolor[rgb]{0.73,0.13,0.13}{##1}}}
\@namedef{PY@tok@si}{\let\PY@bf=\textbf\def\PY@tc##1{\textcolor[rgb]{0.73,0.40,0.53}{##1}}}
\@namedef{PY@tok@se}{\let\PY@bf=\textbf\def\PY@tc##1{\textcolor[rgb]{0.73,0.40,0.13}{##1}}}
\@namedef{PY@tok@sr}{\def\PY@tc##1{\textcolor[rgb]{0.73,0.40,0.53}{##1}}}
\@namedef{PY@tok@ss}{\def\PY@tc##1{\textcolor[rgb]{0.10,0.09,0.49}{##1}}}
\@namedef{PY@tok@sx}{\def\PY@tc##1{\textcolor[rgb]{0.00,0.50,0.00}{##1}}}
\@namedef{PY@tok@m}{\def\PY@tc##1{\textcolor[rgb]{0.40,0.40,0.40}{##1}}}
\@namedef{PY@tok@gh}{\let\PY@bf=\textbf\def\PY@tc##1{\textcolor[rgb]{0.00,0.00,0.50}{##1}}}
\@namedef{PY@tok@gu}{\let\PY@bf=\textbf\def\PY@tc##1{\textcolor[rgb]{0.50,0.00,0.50}{##1}}}
\@namedef{PY@tok@gd}{\def\PY@tc##1{\textcolor[rgb]{0.63,0.00,0.00}{##1}}}
\@namedef{PY@tok@gi}{\def\PY@tc##1{\textcolor[rgb]{0.00,0.63,0.00}{##1}}}
\@namedef{PY@tok@gr}{\def\PY@tc##1{\textcolor[rgb]{1.00,0.00,0.00}{##1}}}
\@namedef{PY@tok@ge}{\let\PY@it=\textit}
\@namedef{PY@tok@gs}{\let\PY@bf=\textbf}
\@namedef{PY@tok@gp}{\let\PY@bf=\textbf\def\PY@tc##1{\textcolor[rgb]{0.00,0.00,0.50}{##1}}}
\@namedef{PY@tok@go}{\def\PY@tc##1{\textcolor[rgb]{0.53,0.53,0.53}{##1}}}
\@namedef{PY@tok@gt}{\def\PY@tc##1{\textcolor[rgb]{0.00,0.27,0.87}{##1}}}
\@namedef{PY@tok@err}{\def\PY@bc##1{{\setlength{\fboxsep}{\string -\fboxrule}\fcolorbox[rgb]{1.00,0.00,0.00}{1,1,1}{\strut ##1}}}}
\@namedef{PY@tok@kc}{\let\PY@bf=\textbf\def\PY@tc##1{\textcolor[rgb]{0.00,0.50,0.00}{##1}}}
\@namedef{PY@tok@kd}{\let\PY@bf=\textbf\def\PY@tc##1{\textcolor[rgb]{0.00,0.50,0.00}{##1}}}
\@namedef{PY@tok@kn}{\let\PY@bf=\textbf\def\PY@tc##1{\textcolor[rgb]{0.00,0.50,0.00}{##1}}}
\@namedef{PY@tok@kr}{\let\PY@bf=\textbf\def\PY@tc##1{\textcolor[rgb]{0.00,0.50,0.00}{##1}}}
\@namedef{PY@tok@bp}{\def\PY@tc##1{\textcolor[rgb]{0.00,0.50,0.00}{##1}}}
\@namedef{PY@tok@fm}{\def\PY@tc##1{\textcolor[rgb]{0.00,0.00,1.00}{##1}}}
\@namedef{PY@tok@vc}{\def\PY@tc##1{\textcolor[rgb]{0.10,0.09,0.49}{##1}}}
\@namedef{PY@tok@vg}{\def\PY@tc##1{\textcolor[rgb]{0.10,0.09,0.49}{##1}}}
\@namedef{PY@tok@vi}{\def\PY@tc##1{\textcolor[rgb]{0.10,0.09,0.49}{##1}}}
\@namedef{PY@tok@vm}{\def\PY@tc##1{\textcolor[rgb]{0.10,0.09,0.49}{##1}}}
\@namedef{PY@tok@sa}{\def\PY@tc##1{\textcolor[rgb]{0.73,0.13,0.13}{##1}}}
\@namedef{PY@tok@sb}{\def\PY@tc##1{\textcolor[rgb]{0.73,0.13,0.13}{##1}}}
\@namedef{PY@tok@sc}{\def\PY@tc##1{\textcolor[rgb]{0.73,0.13,0.13}{##1}}}
\@namedef{PY@tok@dl}{\def\PY@tc##1{\textcolor[rgb]{0.73,0.13,0.13}{##1}}}
\@namedef{PY@tok@s2}{\def\PY@tc##1{\textcolor[rgb]{0.73,0.13,0.13}{##1}}}
\@namedef{PY@tok@sh}{\def\PY@tc##1{\textcolor[rgb]{0.73,0.13,0.13}{##1}}}
\@namedef{PY@tok@s1}{\def\PY@tc##1{\textcolor[rgb]{0.73,0.13,0.13}{##1}}}
\@namedef{PY@tok@mb}{\def\PY@tc##1{\textcolor[rgb]{0.40,0.40,0.40}{##1}}}
\@namedef{PY@tok@mf}{\def\PY@tc##1{\textcolor[rgb]{0.40,0.40,0.40}{##1}}}
\@namedef{PY@tok@mh}{\def\PY@tc##1{\textcolor[rgb]{0.40,0.40,0.40}{##1}}}
\@namedef{PY@tok@mi}{\def\PY@tc##1{\textcolor[rgb]{0.40,0.40,0.40}{##1}}}
\@namedef{PY@tok@il}{\def\PY@tc##1{\textcolor[rgb]{0.40,0.40,0.40}{##1}}}
\@namedef{PY@tok@mo}{\def\PY@tc##1{\textcolor[rgb]{0.40,0.40,0.40}{##1}}}
\@namedef{PY@tok@ch}{\let\PY@it=\textit\def\PY@tc##1{\textcolor[rgb]{0.25,0.50,0.50}{##1}}}
\@namedef{PY@tok@cm}{\let\PY@it=\textit\def\PY@tc##1{\textcolor[rgb]{0.25,0.50,0.50}{##1}}}
\@namedef{PY@tok@cpf}{\let\PY@it=\textit\def\PY@tc##1{\textcolor[rgb]{0.25,0.50,0.50}{##1}}}
\@namedef{PY@tok@c1}{\let\PY@it=\textit\def\PY@tc##1{\textcolor[rgb]{0.25,0.50,0.50}{##1}}}
\@namedef{PY@tok@cs}{\let\PY@it=\textit\def\PY@tc##1{\textcolor[rgb]{0.25,0.50,0.50}{##1}}}

\makeatother

    \makeatletter
        \newbox\Wrappedcontinuationbox 
        \newbox\Wrappedvisiblespacebox 
        \newcommand*\Wrappedvisiblespace {\textcolor{red}{\textvisiblespace}} 
        \newcommand*\Wrappedcontinuationsymbol {\textcolor{red}{\llap{\tiny$\m@th\hookrightarrow$}}} 
        \newcommand*\Wrappedcontinuationindent {3ex } 
        \newcommand*\Wrappedafterbreak {\kern\Wrappedcontinuationindent\copy\Wrappedcontinuationbox} 
        \newcommand*\Wrappedbreaksatspecials {%
            \def\PYGZus{\discretionary{\char`\_}{\Wrappedafterbreak}{\char`\_}}%
            \def\PYGZob{\discretionary{}{\Wrappedafterbreak\char`\{}{\char`\{}}%
            \def\PYGZcb{\discretionary{\char`\}}{\Wrappedafterbreak}{\char`\}}}%
            \def\PYGZca{\discretionary{\char`\^}{\Wrappedafterbreak}{\char`\^}}%
            \def\PYGZam{\discretionary{\char`\&}{\Wrappedafterbreak}{\char`\&}}%
            \def\PYGZlt{\discretionary{}{\Wrappedafterbreak\char`\<}{\char`\<}}%
            \def\PYGZgt{\discretionary{\char`\>}{\Wrappedafterbreak}{\char`\>}}%
            \def\PYGZsh{\discretionary{}{\Wrappedafterbreak\char`\#}{\char`\#}}%
            \def\PYGZpc{\discretionary{}{\Wrappedafterbreak\char`\%}{\char`\%}}%
            \def\PYGZdl{\discretionary{}{\Wrappedafterbreak\char`\$}{\char`\$}}%
            \def\PYGZhy{\discretionary{\char`\-}{\Wrappedafterbreak}{\char`\-}}%
            \def\PYGZsq{\discretionary{}{\Wrappedafterbreak\textquotesingle}{\textquotesingle}}%
            \def\PYGZdq{\discretionary{}{\Wrappedafterbreak\char`\"}{\char`\"}}%
            \def\PYGZti{\discretionary{\char`\~}{\Wrappedafterbreak}{\char`\~}}%
        } 
        \newcommand*\Wrappedbreaksatpunct {%
            \lccode`\~`\.\lowercase{\def~}{\discretionary{\hbox{\char`\.}}{\Wrappedafterbreak}{\hbox{\char`\.}}}%
            \lccode`\~`\,\lowercase{\def~}{\discretionary{\hbox{\char`\,}}{\Wrappedafterbreak}{\hbox{\char`\,}}}%
            \lccode`\~`\;\lowercase{\def~}{\discretionary{\hbox{\char`\;}}{\Wrappedafterbreak}{\hbox{\char`\;}}}%
            \lccode`\~`\:\lowercase{\def~}{\discretionary{\hbox{\char`\:}}{\Wrappedafterbreak}{\hbox{\char`\:}}}%
            \lccode`\~`\?\lowercase{\def~}{\discretionary{\hbox{\char`\?}}{\Wrappedafterbreak}{\hbox{\char`\?}}}%
            \lccode`\~`\!\lowercase{\def~}{\discretionary{\hbox{\char`\!}}{\Wrappedafterbreak}{\hbox{\char`\!}}}%
            \lccode`\~`\/\lowercase{\def~}{\discretionary{\hbox{\char`\/}}{\Wrappedafterbreak}{\hbox{\char`\/}}}%
            \catcode`\.\active
            \catcode`\,\active 
            \catcode`\;\active
            \catcode`\:\active
            \catcode`\?\active
            \catcode`\!\active
            \catcode`\/\active 
            \lccode`\~`\~ 	
        }
    \makeatother

    \let\OriginalVerbatim=\Verbatim
    \makeatletter
    \renewcommand{\Verbatim}[1][1]{%
        \sbox\Wrappedcontinuationbox {\Wrappedcontinuationsymbol}%
        \sbox\Wrappedvisiblespacebox {\FV@SetupFont\Wrappedvisiblespace}%
        \def\FancyVerbFormatLine ##1{\hsize\linewidth
            \vtop{\raggedright\hyphenpenalty\z@\exhyphenpenalty\z@
                \doublehyphendemerits\z@\finalhyphendemerits\z@
                \strut ##1\strut}%
        }%
        \def\FV@Space {%
            \nobreak\hskip\z@ plus\fontdimen3\font minus\fontdimen4\font
            \discretionary{\copy\Wrappedvisiblespacebox}{\Wrappedafterbreak}
            {\kern\fontdimen2\font}%
        }%
        
        \Wrappedbreaksatspecials
        \OriginalVerbatim[#1,codes*=\Wrappedbreaksatpunct]%
    }
    \makeatother

    \definecolor{incolor}{HTML}{303F9F}
    \definecolor{outcolor}{HTML}{D84315}
    \definecolor{cellborder}{HTML}{CFCFCF}
    \definecolor{cellbackground}{HTML}{F7F7F7}
    
    \makeatletter
    \newcommand{\boxspacing}{\kern\kvtcb@left@rule\kern\kvtcb@boxsep}
    \makeatother

    \hypersetup{
      breaklinks=true,  
      colorlinks=true,
      urlcolor=urlcolor,
      linkcolor=linkcolor,
      citecolor=citecolor,
      }
    
    \title{Controlling Chaos in Van Der Pol Dynamics Using Signal-Encoded Deep Learning}
    \vspace{15pt}

    \author{\Large {\sf Hanfeng Zhai} {\em \&}
{\sf Timothy Sands\thanks{To whom correspondence should be addressed. Email: \href{mailto:tas297@cornell.edu}{\tt tas297@cornell.edu}}}
\\\\\emph{{Sibley School of Mechanical
and Aerospace Engineering}},\\
{\it Cornell University, Ithaca, NY}
\\}

\begin{document}
    
    \maketitle
\begin{abstract}
Controlling nonlinear dynamics is a long-standing problem in engineering. Harnessing known physical information to accelerate or constrain stochastic learning pursues a new paradigm of scientific machine learning. By linearizing nonlinear systems, traditional control methods cannot learn nonlinear features from chaotic data for use in control. Here, we introduce Physics-Informed Deep Operator Control (PIDOC), and by encoding the control signal and initial position into the losses of a physics-informed neural network (PINN), the nonlinear system is forced to exhibit the desired trajectory given the control signal. PIDOC receives signals as physics commands and learns from the chaotic data output from the nonlinear van der Pol system, where the output of the PINN is the control. Applied to a benchmark problem, PIDOC successfully implements control with a higher stochasticity for higher-order terms. PIDOC has also been proven to be capable of converging to different desired trajectories based on case studies. Initial positions slightly affect the control accuracy at the beginning stage yet do not change the overall control quality. For highly nonlinear systems, PIDOC is not able to execute control with a high accuracy compared with the benchmark problem. The depth and width of the neural network structure do not greatly change the convergence of PIDOC based on case studies of van der Pol systems with low and high nonlinearities. Surprisingly, enlarging the control signal does not help to improve the control quality. The proposed framework can potentially be applied to many nonlinear systems for nonlinear controls.
    
 \vspace{10pt}
 
\end{abstract}

\section{Introduction\label{intro}}

\lettrine{C}{ontrolling} nonlinear dynamics and chaos is a long-standing issue in various engineering disciplines including aerospace systems design \cite{chaos-aero}, chemical operations \cite{chaos-chem}, robotics \cite{chaos-robo}, biological sciences \cite{chaos-bio}, mechatronics \cite{chaos-mecha}, and specifically, microelectronics \cite{chaos-elec}, especially for circuits systems involving semiconductor that elicit nonlinearity for signal controls \cite{chaos-semiconductor1, chaos-semiconductor2}. In mathematics, chaos is defined as a scenario that typical solutions of a differential equation (or a representation of the system) do not converge to a stationary or periodic function of time but continue to exhibit a seemingly unpredictable behavior \cite{chaos}. Controlling chaos has a simple objective: implementing the desired command to the system to make the system behave "as we wish" or at least to make the system predictable so human impotence can be executed. However, controlling chaos is arduous \cite{chaos-control}, yet ubiquitous in nature as in fluid flows, heartbeat irregularities, weather, and climate \cite{chaos-nat_fluid, chaos-nat_heart, chaos-nat_climate}. Hence, solving such a problem is crucial.

Traditionally, proportional–integral–derivative controllers (PID) were widely adopted for controlling nonlinear systems \cite{pid_1, pid_2}, which employ closed-loop feedback errors that can tune feedforward command as close to the target output as possible\cite{slotine}, are still common in nowadays's industry. Particularly, a common practice in control theory is to implement trigonometric command as for predictable periodic system behavior \cite{tri}. Notably, with such wide applications, PID control experiences sluggish and systematic postpone due to the integral term, and order increasing might lead to system instability \cite{pid_limitation}. Admittedly, reducing PID to PI (proportional-integral) or P (proportional) might either increase speed or system stability, but neither can learn the features of nonlinear systematic data, which allows more advanced self-adjust control behavior.

\begin{figure}[htbp]
    \centering
    \includegraphics[scale=0.42]{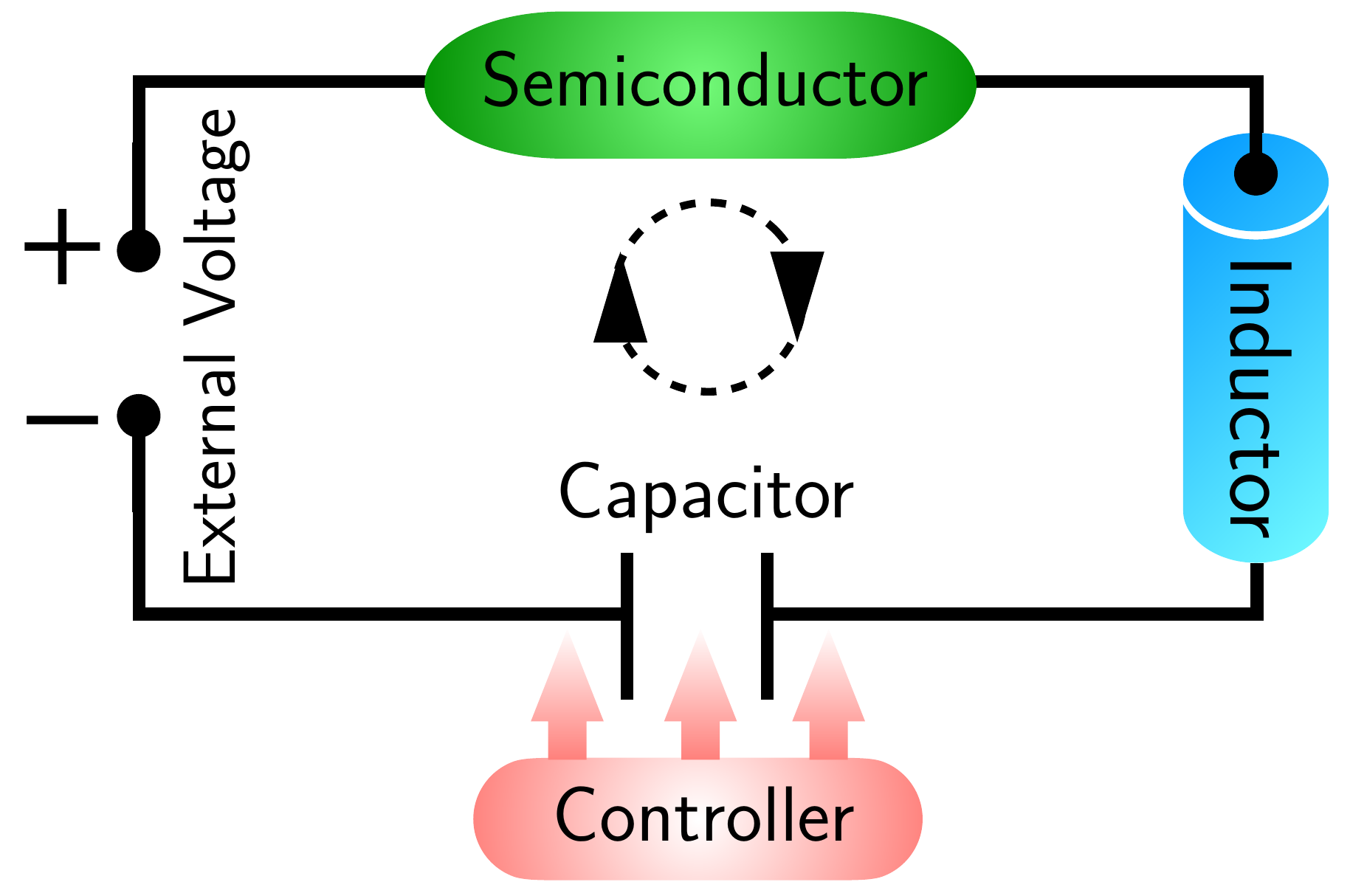}
    \caption{The schematic diagram for the van der Pol circuits. Note that the nonlinearity origins from the semiconductor (the green part), cause the system to generate chaotic behavior. The cycle of the oscillating circuits generates current as illustrated by the dashed circle. See text for details.}
    \label{circuits_schematic}
\end{figure}

In the past decades, the skyrocketing big data, assisted with advanced computing technologies such as GPU computing \cite{gputech}, incited the development of machine learning algorithms, specifically, deep neural networks \cite{deeplearning}, as they can learn and capture features from highly nonlinear data for accurate predictions, showing huge potentials that attracted public's attention in various fields. In recent years, utilizing physics information to encode in the losses of a deep neural network (NN), promising faster accurate learning of physics with NN that respect basic physics laws with less labeled data, commonly recognized as Physics-Informed Neural Networks (PINNs) \cite{pinn, george}. One of the celebrated characteristics of PINNs is they can learn from sparse data \cite{pinn-sparse} as physics doesn't generate humongous data as easily in other commercial fields. Upon the proposed PINN framework, various types of PINNs designed for different engineering applications emerge in fields, with their most renowned works in predicting fluid fields \cite{pinn_science, pinn_turbulence}, but also include electronics \cite{pinn_electric1, pinn_electric2}. Henceforth a question arose: can PINN be applied for dynamical systems? Notably, there are a few attempts using the PINN-based method to learn and predict nonlinear dynamical systems and chaos \cite{pinn_dynamics1, pinn_dynamics2, pinn_nonlinearchaos, pinn_dynamics5}; with a notable good attempt on using it to learn and map the van der Pol system \cite{pinn_pinc}. Notwithstanding, the applications mostly focus on "learn" and "discovery" of dynamics with PINNs, with few actually focusing on "controlling" the system, that is to guide the system to behave as human-desired signals. Also, a few good attempts address NN for controls \cite{nn_control1, nn_control2, nn_control3}, as limited to replacing a block or parts of the closed-loop framework with a NN, rather than directly applying NN-based framework for signal controls.

Acknowledging the limitations of PINNs and other NN-based controls, a question thence emanate: can control signals be incorporated in PINNs as for nonlinear dynamical systems? To investigate such, we focus on a van der Pol system, proposed by van der Pol in the early 20th century while studying oscillating circuits \cite{vdp_1, vdp_2, vdp_3}. The van der Pol system exhibits highly chaotic behavior that encompasses wide applications in biology, biochemistry, and microelectronics \cite{vdp_app1, vdp_app2, vdp_app3}. Traditional control on the van der Pol system usually includes linearizing the system or adding a forced term to impose control \cite{vdp_control1, vdp_control2, vdp_control3, cooper}. A basic setup of the oscillating van der Pol circuit is illustrated in Figure \ref{circuits_schematic}: an external voltage excites the circuit that induces a current, which can be converted to a charge through the capacitor \cite{duke}. If the semiconductor as indicated in green is equipped, the circuit will display highly nonlinear behavior as hard to control and predict. One can design a controller as indicated red block for controlling chaos in circuits, is our main goal.

Inspired by PINNs \cite{pinn}, and focusing on the van der Pol system as different control methods strives to control \cite{cooper}, we proposed Physics-Informed Deep Operator Control (PIDOC), a PINN-based control method that incorporates the generation losses of NN, the desired control signal, and the initial position of the system into the loss function of a PINN that are able to output the desired control signal. The framework is tested based on its behavior and its ability to deal with the highly nonlinear system; the hyperparameters are also investigated for better interpretation and potential applications of PIDOC.

The paper is arranged as follows: in Section \ref{sec_problem} we first formulate the problem of a van der Pol system and controls (Sec. \ref{sec_vdP}), and elaborate the basic system setup in Section \ref{sec_prepare}. In Section \ref{sec_method} the detailed methodologies of formulating and the learning of PIDOC is described, consisting basis of deep learning (Sec. \ref{sec_deeplearn}) and physics-informed control (Sec. \ref{sec_PIcontrol}). It is briefly summarized on how the numerical experiments are conducted in Section \ref{sec_numexp}. Followed by Section \ref{sec_result} showing the results and discussion on PIDOC: Section \ref{sec_systembehavior} analyze the behavior of PIDOC given a benchmark problem; followed by an in-depth estimation of nonlinearity of trajectory convergence in Section \ref{sec_converge}, estimated the amplitude of control signals (Sec. \ref{sec_amplitude}), influences of initial positions (Sec. \ref{sec_ICs}), and noninearity analysis (Sec. \ref{sec_nonlinearity}). The hyperparameters of the NN (Sec. \ref{sec_deepnn}) and the weight of the control signal in PINN (Sec. \ref{sec_lagran}) is further studied in Section \ref{sec_hyperparam}. The paper is eventually concluded and summarized in Section \ref{sec_conclusion}.

\section{Problem Formulation\label{sec_problem}}

The nonlinear control of the chaotic van der Pol system problem is mainly inspired by the work of Cooper {\em et al.} \cite{cooper}, with a clear problem objective: the effort of a successful implementation of the desired trajectory to the van der Pol system.

\subsection{van der Pol dynamics\label{sec_vdP}}

van der Pol describes the oscillatory behavior to the class of nonlinear equations that are now referred to the van der Pol equation \cite{vdp_1}, where originally proposed to describe oscillating circuits with semiconductors as introduced in Section \ref{intro}, writes: \begin{equation}
    \frac{d^2 x}{dt^2} - \mu (1 - x^2) \frac{dx}{dt} + x = 0\label{vdpeq}
\end{equation}where if $x(t)$ is referred as position, thence  $\dot{x}(t)$ is the velocity and $\ddot{x}(t)$ is the acceleration; $\mu$ is a scalar parameter indicating the nonlinearity and the strength of the damping.

\begin{figure}[htbp]
    \centering
    \includegraphics[scale=0.55]{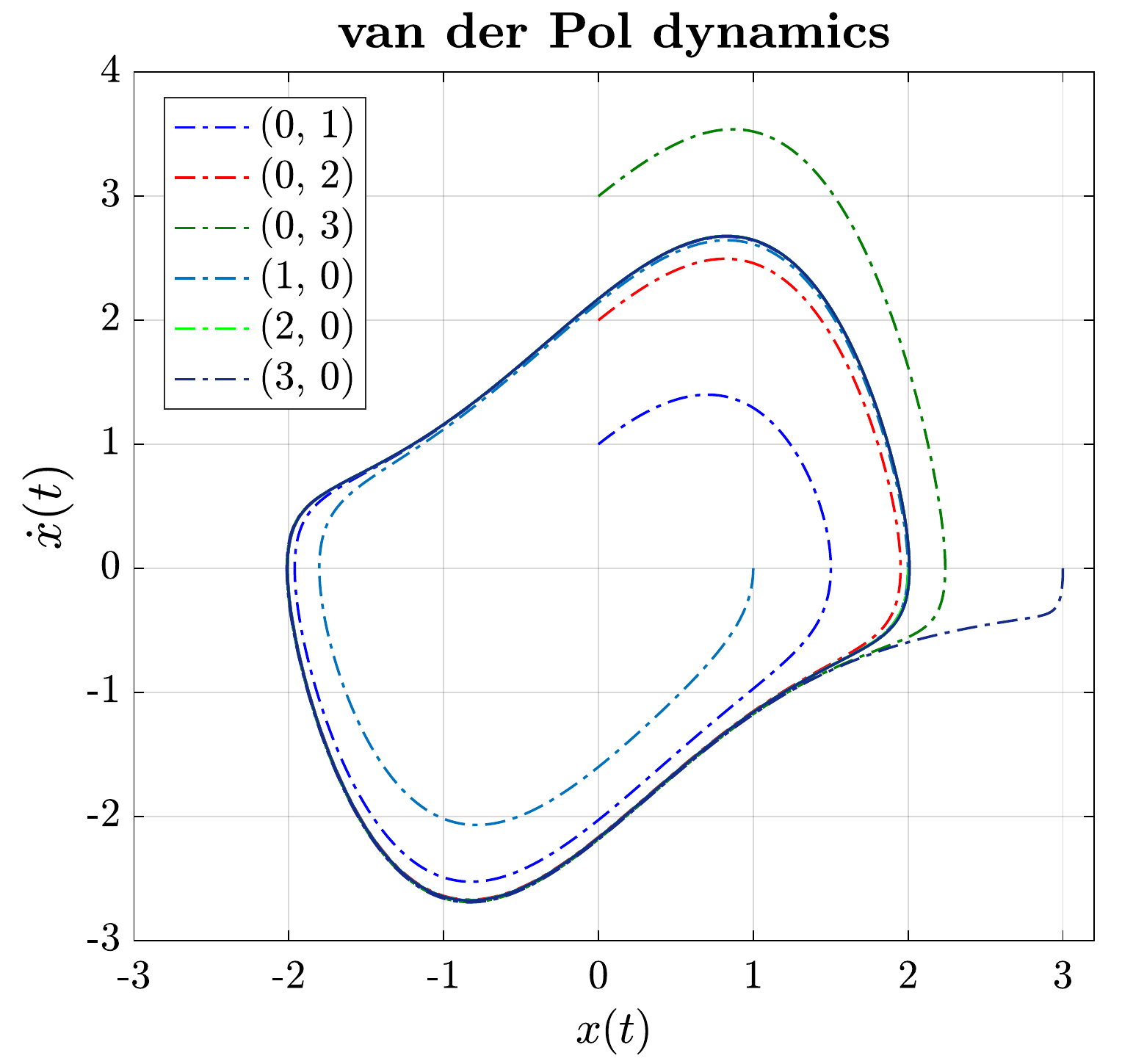}
    \caption{The phase portrait for the nonlinear van der Pol equation when $\mu = 1$. The phase is initiated from six different points: $(1, 0)$, $(2, 0)$, $(3, 0)$, $(0, 1)$, $(0, 2)$, $(0, 3)$, as indited in different dashed lines in the box, both converge to the same nonlinear trajectory of the van der Pol inherent dynamics. Note that the data for this figure is generated using \texttt{odeint} in \texttt{SciPy} library.}
    \label{vdP_dynamics}
\end{figure}

The equation exhibits an oscillatory behavior, yet with a non-constant amplitude, representing an invariant trajectory set called a “limit cycle” \cite{cooper}. System trajectories converge to these invariant orbits given any initial conditions, as can be referred to in Figure \ref{vdP_dynamics}: Given six different picked initial points, all converge to the non-constant limit cycle, with a changing velocity to positions, indicating a non-stable phase. The robustness of the nonlinear phase suffers scientists as the system is automatically "trapped" in the state as long as the control signals cease, indicating the robustness of the inherent dynamics.

As indicated in Section \ref{intro}, the main goal addressed is to control such nonlinear behavior. Seeking to produce a fixed-amplitude oscillation, traditional control methods commonly add forcing functions to the nonlinear equation as in Equation (\ref{forced}), which allows linear time-invariant (LTI) feedback controller based on a linearized version of the system equation, as for classical control design \cite{cooper}:
\begin{equation}
    \frac{d^2 x}{dt^2} - \mu (1 - x^2) \frac{dx}{dt} + x = F (t)\label{forced}
\end{equation}
However, in this paper such an attempt is not adopted: the goal is to control the unsteady state of oscillating dynamics by learning from the original van der Pol equation receiving an external control signal to the controller, with no external terms added to modify the original van der Pol formulation. Such an attempt won't suffer from the errors of linearization as in traditional methods such as PI or PID controllers \cite{pi, pid_1, pid_2}, neither does it accumulate the errors from integration as in PID controls. What's more, the NN-based control PIDOC is able to capture the feature from nonlinear chaotic data directly from the van der Pol system (through the capacitor in Figure \ref{circuits_schematic}); and the optimization of minimizing losses is an innate feedback process to the framework for control as the errors are reduced per iterations, with the feed-forward control signal is output as the predictions from the learning of the NN. For system controls, Input a sinusoidal or other triangular functions through modifying $F(t)$ in Equation (\ref{forced}) is a classical approach; where we incorporate such a signal as an external supervision \cite{bubble} added to PIDOC in the paper.

\subsection{System simulation\label{sec_prepare}}

Depicting and emulate the system as described in Figure \ref{circuits_schematic} requires a simulation of van der Pol system that generate nonlinear data, which are fed into PIDOC for control. In the classical definition of a NN, such data are considered as training data, for the NN to "learn" and "predict". Here, the \texttt{Python} differential equation solver \texttt{odeint}, part of \texttt{SciPy} library, is adopted to solve van der Pol equation for system simulation \cite{vdpsolver}. The \texttt{odeint} is a library containing advanced numerical methods for solving differential equations, especially for initial value problems \cite{odeint}. Compare with other solver such as \texttt{scipy.integrate.solve\_ivp}, \texttt{odeint} are able to generate data with higher smoothabilities \cite{solver_compare}, promising a "cleaner" data fed for NN training and learning. In this paper, the van der Pol equation is solved for $t = 30$, interpolated with 3000 points, as in Equation (\ref{vdpeq}). The parameters determine the error control performed by the solver \texttt{rtol} and \texttt{atol} is chosen as $10^{-6}$ and $10^{-10}$, respectively \cite{vdpsolver}.

\section{Methodology and Algorithm\label{sec_method}}

In this section, we will introduce the basic setup of PIDOC, employing PINN as initial position (denoted by $\mathcal{I}$) and control signal (denoted by $\mathcal{D}$, meaning desired trajectory) encoded within the NN losses for accurate controls.

\subsection{Deep learning\label{sec_deeplearn}}

PIDOC controls nonlinear systems based on "learning" from chaotic data for output a prediction as control, which are enabled by a deep neural network (DNN). For the van der Pol system, the DNN is formulated as:\begin{equation}
    \begin{aligned}
    x_{pred} = (K_L \circ \sigma_L \circ ... \circ K_1 \circ \sigma_1 \circ K_0) t\label{nn}
    \end{aligned}
\end{equation}where the DNN outputs an desired trajectory $x_{pred}$ given an input of the specific time series $t$. $K_1, K_2, ..., K_L$, are linear layers; $\sigma_1, \sigma_2, ..., \sigma_L$ are the activation functions, where PIDOC employs \texttt{tanh} activation functions. For PIDOC, the NN take time $t$ as the input layer $K_0$ to transmit through $(L-1)^{th}$ hidden layers to generate an output $x_{pred}$ through the output layer $K_L$, supervised from the chaotic data of the van der Pol system $x_{train}$. Here, Let $\mathcal{N}^L \equiv (K_L \circ \sigma_L \circ ... \circ K_1 \circ \sigma_1 \circ K_0)$ denotes the $L^{th}$ layers NN.

\begin{figure}[htbp]
    \centering
    \includegraphics{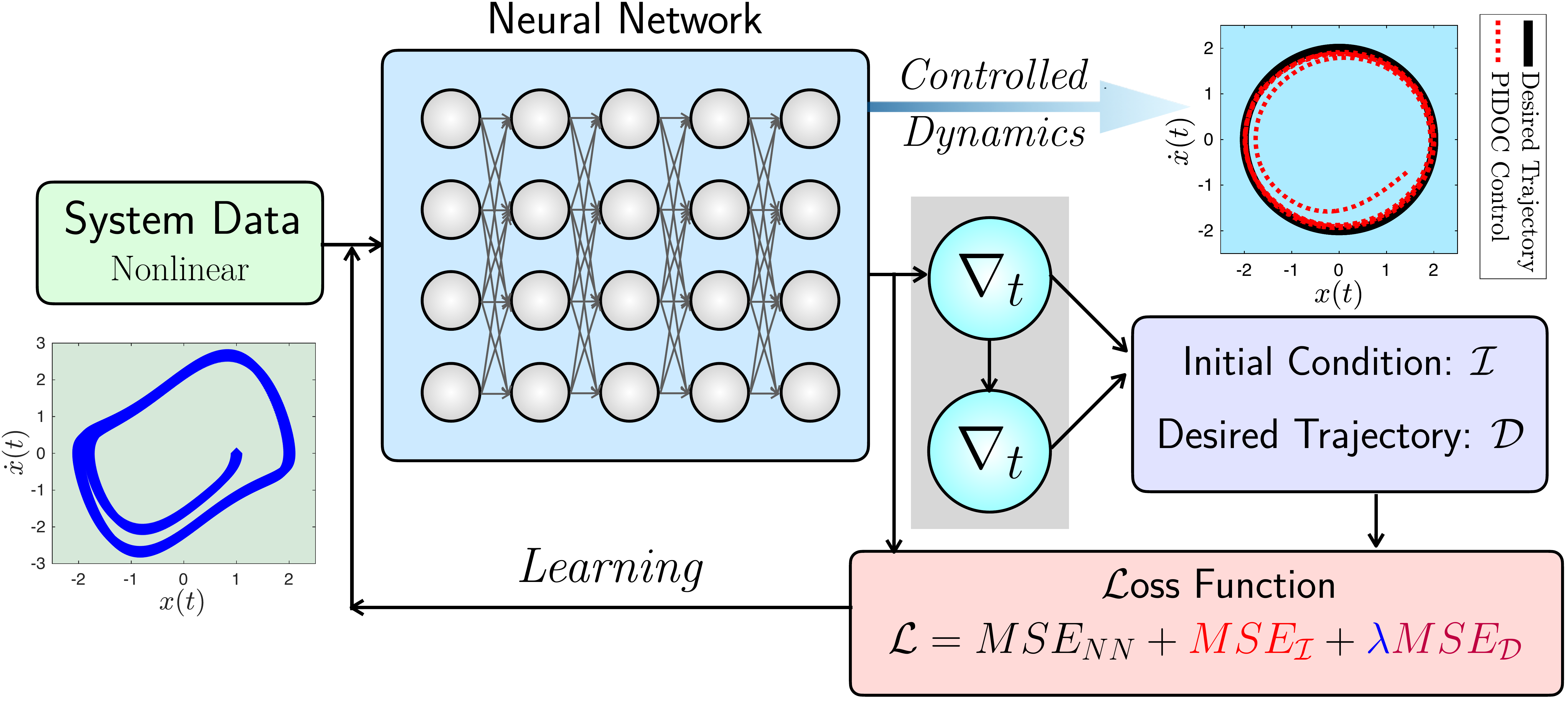}
    \caption{The schematic diagram for \textsc{Physics-Informed Deep Operator Control} framework was inspired by PINN. The system data of the nonlinear van der Pol oscillator is input to the deep learning framework, as automatic differentiation can encode physics information as in the purple box. The encoded information is further forward to the loss function, as the Langrangian multiplier $\lambda$ can enlarge the control signal, which feedbacks to the deep learning scheme for learning the dynamics, which can output the ideal dynamics.}
    \label{PIDOC_schematic}
\end{figure}

The DNN corresponds to Equation (\ref{nn}) are commonly recognized as a combination of three parts: input, hidden, and output layers, where the neurons are connected, commonly known as a feed-forward NN, defined recursively as\begin{equation}
    \begin{aligned}
    {\rm Input\ layer}:\ K_0 (t) &= t \subset \mathbb{R}^{d_{in}}\\
    {\rm Hidden\ layer}:\ K_\mathfrak{L} (t) &= \sigma \left( {\bf w}_\mathfrak{L} K_{\mathfrak{L}-1} (t) + {\bf b}_\mathfrak{L}\right)\subset \mathbb{R}^{\mathcal{K}^\mathfrak{L}},\ {\rm for}\ 1 \leq \mathfrak{L} \leq L-1\\
    {\rm Output\ layer}:\ K_L (t) &= {\bf w}_LK_{L-1} + {\bf b}_L \subset \mathbb{R}^{d_{out}},\ \left(x_{pred} \equiv K_L (t)\right)\label{nn_lu}
    \end{aligned}
\end{equation}For the NN $\mathcal{N}^L(t): \mathbb{R}^{d_{in}}\longrightarrow\mathbb{R}^{d_{out}} $,  denotes the input time series $t$ is transmit to the linear input layer $K_0$; forward through hidden layers, from a linear model with $\mathbf{w}_\mathfrak{L}$ as weights and $\mathbf{b}_\mathfrak{L}$ as biases activated through activation function \texttt{tanh}, to generate an output through $K_L$ for NN predictions $x_{pred}$, can also interpreted as the {\em controlled dynamics}. To be videlicet, there are $\mathcal{K}^\mathfrak{L}$ neurons in the $\mathfrak{L}^{th}$ layer $\left( \mathcal{K}^0 = d_{in}\ \&\ \mathcal{K}^L = d_{out}\right)$, the weights and biases are thence denoted by $\mathbf{w}_{\mathfrak{L}} \subset \mathbb{R}^{\mathcal{K}^\mathfrak{L} \times \mathcal{K}^{\mathfrak{L} - 1}}$ and $\mathbf{b}_\mathfrak{L} \subset \mathbb{R}^{\mathcal{K}^{\mathfrak{L}}}$ \cite{deepxde}. Note that the visualization of the NN is shown in the blue box in Figure \ref{PIDOC_schematic}.

In the PIDOC formulation, a supervised machine learning problem is defined, where the NN learning is enabled through the supervision of external training data, as a formulation on minimizing the loss function, so that the NN can capture data structures through this optimization process, where in traditional NN aproaches $\mathcal{L}$ is usually the differences (errors) between the NN predictions and training data. Let $\mathcal{L} = \mathcal{L}(t, {\bf p})$ denotes the loss function, where $t$ is the input time series and $\bf p$ is the parameters vectors containing in formations of $\mathcal{I}$, $\mathcal{D}$, and NN. If no external constraints or bounds are enforced, the optimization problem hence taking the form:\begin{equation}
    \min_{t \subset \mathbb{R}^{d_{out}}} \mathcal{L}(t, {\bf p})\label{opt}
\end{equation}Minimizing $\mathcal{L}$ requires recursive iterations over the NN as in Equation (\ref{nn_lu}): the information encoded in $\mathcal{L}$ reduces during iterations given the optimization methods; where we adopt limited-memory Broyden–Fletcher–Goldfarb–Shanno optimization algorithm, a quasi-Newton method (\textsf{L-BFGS-B} in \texttt{TensorFlow 1.x}) \cite{bfgs, tensorflow}. The optimization are carried over iterations loop from the blue box (NN) to purple box ($\mathcal{I}$ \& $\mathcal{D}$) to red box ($\mathcal{L}$) as in Figure \ref{circuits_schematic}. The maximum iterations are set as $2\times10^5$. In the PIDOC formulation, $\mathcal{L}$ is calculated based on mean square errors of the encoded information to be construed in Section \ref{sec_PIcontrol}.

\subsection{Physics-Informed control\label{sec_PIcontrol}}

The implementation of control signals are enabled through the physics information inserted in the loss function, mimicking the strategy of PINNs, but instead aim to execute command disparate to the training data as to {\em tune} the system into a stable stage. The formulation of the loss function includes the mean square errors (MSE) of the NN generation $MSE_{NN}$, initial conditions $MSE_{\mathcal{I}}$ and the desired trajectory $MSE_\mathcal{D}$ multiplies a Lagrangian multiplier $\lambda$ as to enlarge the control signal:\begin{equation}
    \begin{aligned}
    \mathcal{L} = MSE_{NN} + MSE_{\mathcal{I}} + \lambda MSE_{\mathcal{D}}\label{losses}
    \end{aligned}
\end{equation}where the NN generation errors $MSE_{NN}$ are computed as the MSE between the difference of training data $x_{train}$ and NN predicted output $x_{pred}$:\begin{equation}MSE_{NN} := \frac{1}{N} \sum_{i=1}^N \left|x_{train} - x_{pred}\right|^2 \label{msenn}\end{equation} The initial position loss $MSE_{\mathcal{I}}$ are computed as the MSE between given initial conditions of $\mathcal{D}$ with system predictions $x_{pred}$ at the initial:\begin{equation}MSE_{\mathcal{I}} := \frac{1}{N} \sum_{i=1}^N \left|x_{pred}^0 - x_{\mathcal{D}}^0\right|^2\label{initial_loss}\end{equation}where $x_{pred}^0$ denotes the initial position ($0^{th}$ in the array in \texttt{Python}) of NN prediction; $x_{\mathcal{D}}^0$ denotes the initial position of desired trajectory. 

The control signal losses $MSE_{\mathcal{D}}$ are the MSE between the differences between the zeroth and second order derivatives of the position between the NN predictions (output) and desired control trajectories. In short, $MSE_\mathcal{D}$ imposes our desired control to PIDOC:\begin{equation}MSE_{\mathcal{D}} := \frac{1}{N} \sum_{i=1}^N \left|\left(\frac{d x_{\mathcal{D}}^2}{dt^2} - \frac{d x_{pred}^2}{dt^2}\right) + \left(x_\mathcal{D} - x_{pred}\right)\right|^2\label{desire_loss}\end{equation}Here, $MSE_{\mathcal{D}}$ incorporates the control, and the multiplication of $\lambda$ in Equation (\ref{losses}) allows us to tune the weight of the signal. The optimization problem formulated in Equation (\ref{opt}) can hence be considered as a multi-objective gradient-based optimization, as the objective of control can be tuned through $\lambda$. Specifically, it is reported that the limited memory BFGS method has been wide applied to large scale unconstrained optimizations \cite{nyuphd}.

Classical control approach impose a square wave or a triangular function as signals \cite{wave}. For circuits system specifically, applying a sinusoidal wave is a common practice, as also did by Cooper {\em et al.} \cite{cooper}. Here, for applications of PIDOC, we also applied a sinusoidal wave, multiplies an adjustable multiplier $\Lambda$ to control the amplitude of the phase: \begin{equation}
    x_\mathcal{D}(t) = \Lambda\sin(t),\ \Longrightarrow \dot{x}_\mathcal{D}(t) = \Lambda\cos(t), \ \ddot{x}_\mathcal{D}(t) = -\Lambda\sin(t)\label{amplitude}
\end{equation}Given $x_\mathcal{D}$, the output phase portrait ($x(t)$ - $\dot{x}(t)$ diagram) is expected to be a circular trajectory. However, to make PIDOC adjustable with the desired trajectory amplitude $\Lambda$, one should modify Equation (\ref{msenn}) as to make the NN losses contain information of trajectory amplitude, with the same training data: \begin{equation}MSE_{NN} := \frac{1}{N} \sum_{i=1}^N \left|x_{train} - \frac{x_{pred}}{\Lambda}\right|^2\end{equation}

To perform decent system behavior analyses, we elicit four parameters for comparing behaviors involved in designing and applying PIDOC for control, as follows.

The accuracy of the controlled position (NN predictions) by PIDOC can be quantified by computing the absolute errors mean value between the desired position $x_\mathcal{D}$ with the PIDOC output $x_{pred}$, averaged over the data samples $N$:\begin{equation}
    \begin{aligned}
    \left|\overline{\mathcal{E}}\right| \equiv \left|\overline{\mathcal{E}}_{x(t)}\right|= \left| \frac{1}{N}\sum_{i=1}^N \left(\frac{x_{pred} - x_\mathcal{D}}{x_\mathcal{D}}\right)\right|
    \end{aligned}\label{abserr}
\end{equation}where $N=3000$ in our case of applying PIDOC to van der Pol system.

The average value of the objective function (or loss function in the NN) $\mathcal{L}$ quantifies the accuracy of control (NN prediction) during the optimization process as in Equation (\ref{opt}), nominated as the mean losses $\overline{\mathcal{L}}$, calculated as the losses per iterations\begin{equation}
    \begin{aligned}
    \overline{\mathcal{L}} =  \frac{1}{M}\sum_{i=1}^M \left(\mathcal{L}\right)
    \end{aligned}\label{meanloss}
\end{equation}where $M$ is the iterations number, which are not fixed and dependent on the convergent criteria of \textsf{L-BFGS-B} \cite{bfgs}.

Computing power consumption can be quantified recalling the machine running time (\texttt{timeit.default\_timer()} module in \texttt{Python}), which is symbolized as $\mathcal{T}$ as computation time. Note that the training of the NN is carried out on Google Colab \cite{colab}, hence the unit of the exact time may not accurately reflect the computer platform, yet the quantity differences can qualitatively reflect the system behavior.

Since the computation time are collected over an uncertain iterations number $M$, averaging $\mathcal{T}$ over $M$, and normalized over a benchmark time $\hat{\mathcal{T}}^\dagger$ could proffer a decent quantification of computing time per iterations during the NN learning, called the mean normalized time $\tilde{\mathcal{T}}$: \begin{equation}
    \begin{aligned}
    \tilde{\mathcal{T}} = \frac{\mathcal{T}}{M}\frac{1}{\hat{\mathcal{T}}^\dagger}
    \end{aligned}\label{meantime}
\end{equation}where the computing time of the benchmark problem $\hat{\mathcal{T}}^\dagger$ is averaged through $\hat{\mathcal{T}}^\dagger = \mathcal{T}^\dagger / M^\dagger$, where $\mathcal{T}^\dagger$ is the total computing time and $M^\dagger$ is the iterations of the benchmark problem, to be elaborated in Section \ref{sec_numexp}.

\subsection{Numerical experiments\label{sec_numexp}}

The systematic behavior of PIDOC is analyzed through numerical experiments covering two aspects: the capacity of PIDOC for controlling different systems, and how the hyperparameters and intrinsic structure of PIDOC variate its control process. Initially, we apply a benchmark problem for estimating the control process of PIDOC: the amplitude of the desired trajectory $\Lambda = 2$ in Equation (\ref{amplitude}); the van der Pol system with nonlinear parameter $\mu = 1$ in Equation (\ref{vdpeq}); given an initial point of $(1, 0)$ corresponds to Figure \ref{vdP_dynamics}; taking the Lagrangian multiplier $\lambda = 1$ as in Equation (\ref{losses}); with a neural network of 30 neurons per 6 layers (denoted as $6\times 30$). We first apply PIDOC to control the benchmark problem and analyze its system behavior. Further, for how PIDOC behaves differently for different systems, we first change the trajectory amplitude $\Lambda$ from 1 to 5 and study the PIDOC behavior; followed by changing initial positions as $(1, 0)$, $(5, 0)$ and $(0, 5)$ for studying how initial positions may affect PIDOC controls; followed by changing the nonlinearity on $\mu$ in Equation (\ref{vdpeq}) to check how PIDOC controls different nonlinear systems. Regarding the PIDOC basic architecture and its influence on the control signals, we first change the NN architecture (blue box in Figure \ref{PIDOC_schematic}) and apply PIDOC to the benchmark problem of $\mu = 1$; we further increase the layers as try to apply PIDOC for controlling system with high nonlinearities ($\mu = 5$ in Equation (\ref{vdpeq})). To study whether the enlargement of the control signal can increase or decrease the effectiveness of controls, the Lagrangian multiplier $\lambda$ in Equation (\ref{losses}) is changed for five different values, $0, 1, 10, 10^3, \infty$, to test the PIDOC systematic behavior. Note that for $\lambda = \infty$, we only save the $MSE_\mathcal{D}$ term and eliminate the rests as this numerically represent $\lambda \longrightarrow \infty$. The estimation on van der Pol system and PIDOC architectures are based on the benchmark problem as only changing the parameters to be investigated. To eliminate the errors of reloading Google Colab for calculation of training time and other parameters, the numerical experiments are redone for each table to investigate each factor in each section.

\section{Results and Discussion\label{sec_result}}

\subsection{System behavior analysis\label{sec_systembehavior}}

\begin{figure}[htbp]
    \centering
    \includegraphics{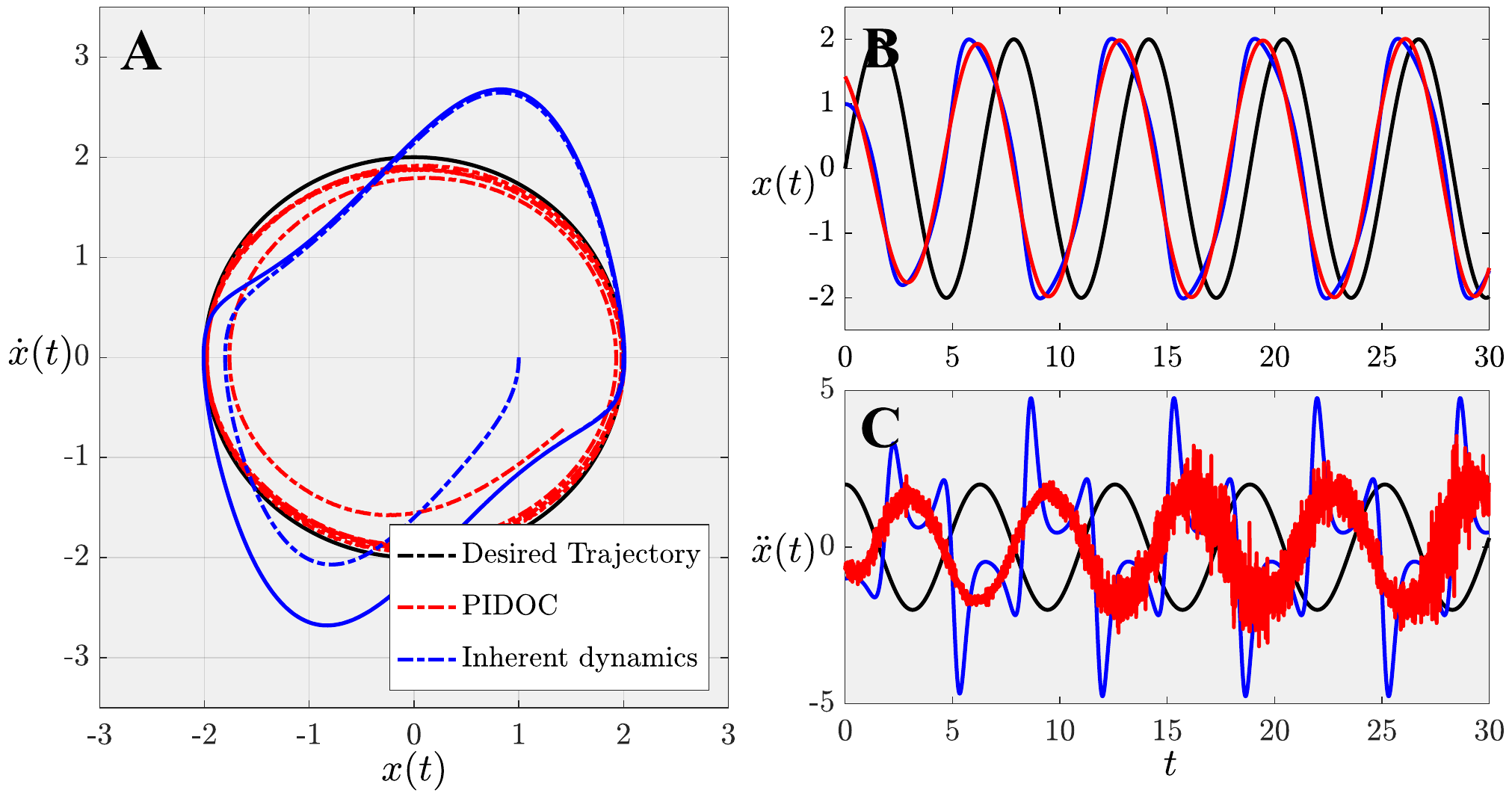}
    \caption{The system behavior of using PIDOC to control van der Pol dynamics applied to the benchmark problem. {\bf A}. the phase portrait of the desired trajectory $\mathcal{D}$, PIDOC controlled signal, and the van der Pol inherent dynamics. {\bf B}. the plot of position $x(t)$ regarding time $t$. {\bf C}. the plot of acceleration $\ddot{x}(t)$ regarding time $t$.}
    \label{systemanalysis}
\end{figure}

The PIDOC framework is first applied to the benchmark problem as introduced in Section \ref{sec_numexp}, the systematic behavior is tested based upon the phase portrait, the time scheme of position $x(t)$ and acceleration $\ddot{x}(t)$, respectively, both shown in Figure \ref{systemanalysis}. From Figure \ref{systemanalysis} {\bf A} in Figure \ref{systemanalysis} it can be detected that PIDOC successfully implement control to the desired trajectory as shown in the red dashed line converging to the circle of $\Lambda = 2$, getting rid of the chaotic behavior of the van der Pol system in blue dashed line. However, it can be noticed in Figure \ref{systemanalysis} {\bf B} that the controlled scheme of PIDOC as in the red line exhibits a phase difference with the desired route as in the black line. It can also be observed in Figure \ref{systemanalysis} {\bf C} the PIDOC controlled scheme displays an oscillating behavior as indicated in the red line.  

Three characteristics of PIDOC control can be concluded from Figure \ref{systemanalysis}: (1) PIDOC successfully implemented the control to guide the system behave based upon $\mathcal{D}$; (2) PIDOC controlled dynamics exhibits a phase lag with the desired signals; (3) the acceleration $\ddot{x}(t)$ exhibits a fluctuating behavior. For (2) and (3), the following explanations are provided: the phase difference in (2) (Figure \ref{systemanalysis} {\bf B}) are attributed to PIDOC aims to follow the given initial positions $\mathcal{I}$ encoded in Equation (\ref{losses}). The stochasticity for (3) observed in Figure \ref{systemanalysis} \textbf{C} can be attributed to both the stochastic nature of NN training and numerical differential of position $x(t)$: the weights ${\bf w}_\mathfrak{L}$ in Equation (\ref{nn_lu}) are randomized and renewed per iterations as to approximate the nonlinear data given for training, as it seems smooth for first order approximation of $x(t)$ in Figure \ref{systemanalysis} {\bf B}, the higher order terms $\ddot{x}(t)$ will enlarge the stochasticity of the signal. Moreover, the formulation of PIDOC as adopted from PINNs \cite{pinn, george, deepxde}, automatic differentiation also elicit errors as encoded in $\mathcal{L}$ (Equation (\ref{losses})) \cite{errorprop}. Both can also be accounted as factors as to explain fluctuations in Figure \ref{systemanalysis} {\bf C}.

\subsection{Nonlinearity and trajectory convergence\label{sec_converge}}

\subsubsection{Amplitude of control trajectory\label{sec_amplitude}}

By changing amplitudes of the desired trajectories, PIDOC is tested on the capacities of executing different control signals, as indicated in Section \ref{sec_numexp}. With the given five trajectories of $\Lambda = 1 \sim 5$, the phase portraits are shown in Figure \ref{amplitude} {\bf A}. The losses corresponding to each trajectory during the NN training are shown in Figure \ref{amplitude} {\bf B}. Figure \ref{amplitude} {\bf A} contends that as trajectory amplitudes $\Lambda$ increases there is a more evident difference between the PIDOC controls and $\mathcal{D}$. However, PIDOC exhibits good signal implementation and gets rid of the inherent dynamics successfully for both amplitudes of desired trajectories. Figure \ref{amplitude} {\bf B} indicates the losses reduced to the lowest value as compared with other $\Lambda$ with the least iterations, as $\Lambda = 2\ \&\ 3$ losses are in the same numerical value, same as $\Lambda = 4\ \&\ 5$, during a value between $10^2$ to $10^3$.

\begin{table}
    \centering
    \begin{tabular}{c|c c c c}\hline
        $\Lambda$ & $|\overline{\mathcal{E}}|$ & $\mathcal{T}$ & $\overline{\mathcal{L}}$ & $\tilde{\mathcal{T}}$\\\hline
        1 & $2.2501\times10^{-4}$ & $1.9554\times10^{4}$ & $2.3630\times10^{3}$ & $1.0000$\\
        2$^\dagger$ & $2.2520\times10^{-4}$ & $1.4098\times10^{4}$ & $1.0562\times10^{3}$ & $0.5642$\\
        3 & $2.2413\times10^{-4}$ & $2.0273\times10^{4}$ & $1.1336\times10^{3}$ & $0.9275$\\
        4 & $2.2330\times10^{-4}$ & $2.0759\times10^{4}$ & $1.4812 \times10^{3}$ & $0.8074$\\
        5 & $2.2241\times10^{-4}$ & $2.1210\times10^{4}$ & $1.6920\times10^{3}$ & $0.8443$\\\hline
    \end{tabular}
    \caption{Parameters estimation of the trajectory amplitude $\Lambda$. $^\dagger$The benchmark setup used for parameter tuning.}
    \label{table_lagrange}
\end{table}

\begin{figure}[htbp]
    \centering
    \includegraphics[scale=1.1]{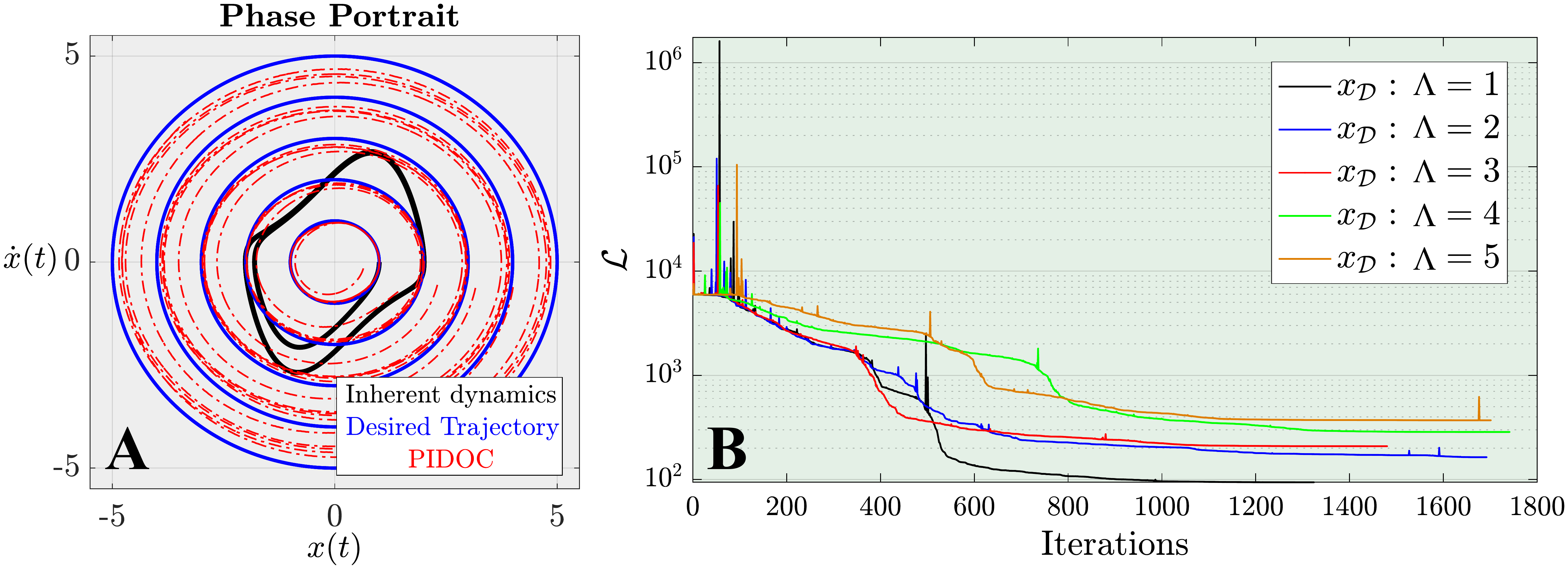}
    \caption{Systematic behavior analysis of PIDOC applied to different trajectory amplitudes $\Lambda$. {\bf A}. the phase portrait of the inherent dynamics, five different desired trajectories $\Lambda = 1,2,..., 5$ marked in blue solid lines and the corresponding PIDOC controlled output marked in red dotted lines. {\bf B}. the loss function - iterations diagram for five trajectories.}
    \label{fig_trajectory}
\end{figure}

The absolute mean errors (Equation (\ref{abserr})), training time, mean losses (Equation (\ref{meanloss})), and the mean normalized time (Equation (\ref{meantime})), corresponding to Figure \ref{fig_trajectory}, shortened as "{PIDOC estimates}", are shown in Table \ref{table_lagrange}, which unveils information that are not straightforward in Figure \ref{fig_trajectory}. The $|\overline{\mathcal{E}}|$ values indicates for targeted trajectories with higher amplitudes, the relative errors are lower. For a higher trajectories' values, the training time $\mathcal{T}$ are higher. The average losses $\overline{\mathcal{L}}$ show good agreement with Figure \ref{fig_trajectory} except for $\Lambda = 1$: an evidently higher mean loss than other trajectories. One can explain such a phenomenon with the instability of weights and biases generation of NN training caused the loss explosion, which is common for NN. For the normalized mean time $\tilde{\mathcal{T}}$, an unanticipated phenomenon of shorter $\tilde{\mathcal{T}}$ with higher amplitudes of trajectories is reported.

\subsubsection{Initial positions\label{sec_ICs}}

\begin{figure}[htbp]
    \centering
    \includegraphics[scale=1.1]{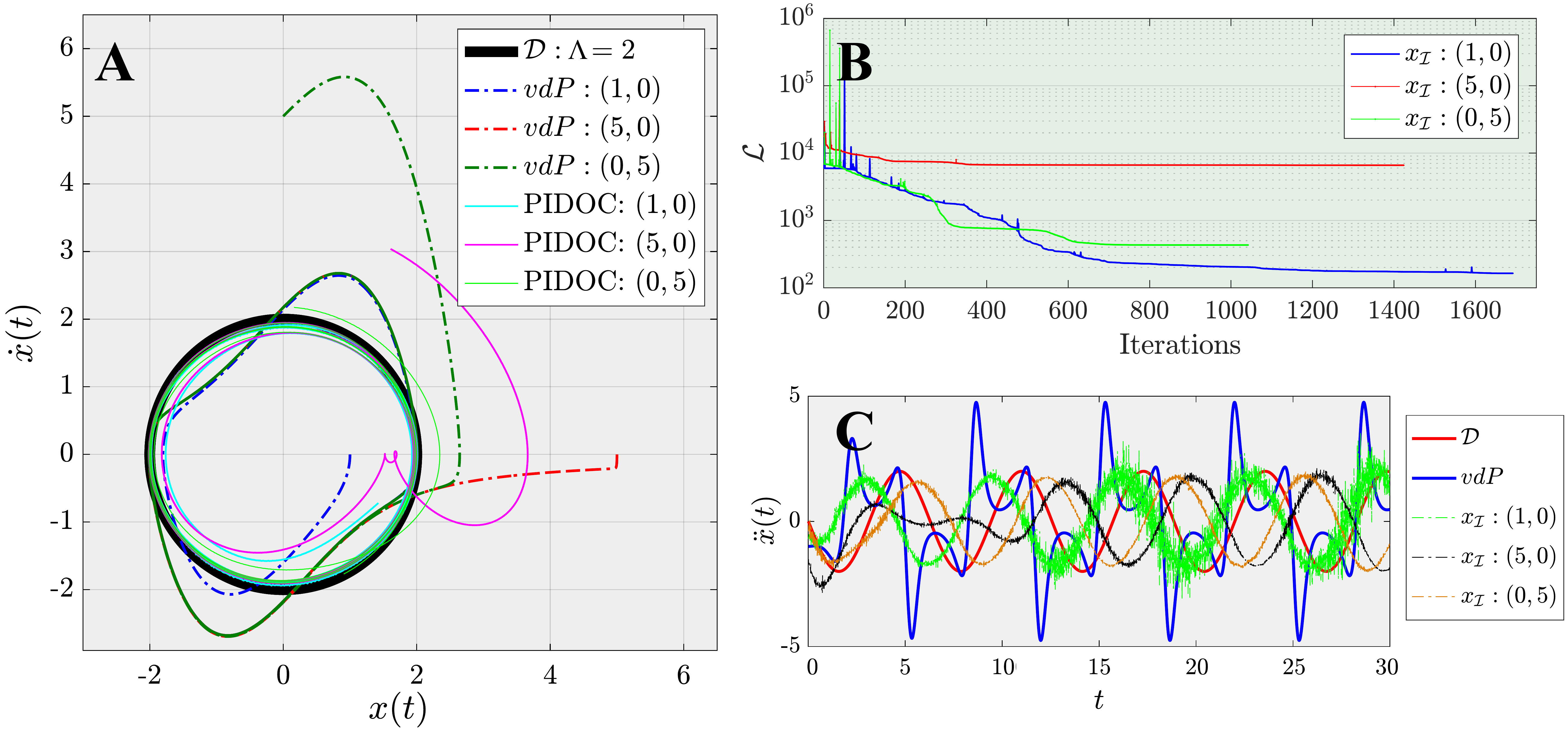}
    \caption{Systematic behavior estimation of PIDOC applied to the control signals $\Lambda = 2$ with different initial positions. Note that $\mathcal{D}$, $vdP$ in plot legend stands for desired trajectory and van der Pol inherent dynamics. {\bf A}. phase portrait of the desired trajectory, inherent dynamics given three different initial points $(1,0)$, $(5,0)$, $(0,5)$; and PIDOC controlled trajectories given three initial positions. {\bf B}. the loss function - iterations diagram is given three different initial points. {\bf C}. the acceleration $\ddot{x}(t)$ - time plot of desired trajectory $\mathcal{D}$, van der Pol inherent dynamics, and the three PIDOC applied controlled routes given different initial positions.}
    \label{fig_initialpoints}
\end{figure}

Given three different initial positions $\mathcal{I}$: $(1, 0)$, $(5, 0)$, $(0, 5)$, the PIDOC control and systematic responses are shown in Figure \ref{fig_initialpoints}. Note that the information of $\mathcal{I}$ are encoded to PIDOC within the physics-informed control in Equations (\ref{losses}), (\ref{initial_loss}). Figure \ref{fig_initialpoints} {\bf A} implies all PIDOC controls with different $\mathcal{I}$s successfully converge to the desired trajectory, with the control $\mathcal{I}$ of $(0, 5)$ exhibits a slightly higher fluctuation as in the green line and the control $\mathcal{I}$ of $(5, 0)$ exhibits a strong trajectory mismatch at the initial stage as indicated in the pink line. The losses in Figure \ref{fig_initialpoints} {\bf B} implies when $\mathcal{I}$ is $(1, 0)$ PIDOC exhibits the lowest loss while for $\mathcal{I}$ is $(5, 0)$ the loss is the highest, which agrees well with Figure \ref{fig_initialpoints} {\bf A}. Figure \ref{fig_initialpoints} {\bf C} shows that PIDOC with $\mathcal{I}$ of $(5, 0)$ and $(0, 5)$ displays a slightly weaker fluctuations as in red and orange dashed lines compared with the green one. Specifically, both PIDOC controls for three $\mathcal{I}$s have an obvious phase difference, yet all converge to the desired trajectory. Recall our estimation for Figure \ref{systemanalysis}, such a phase difference can be attribute to the approximation for initial positions by PIDOC as we encode such an information in Equation \ref{initial_loss}, which decently explains the phase difference in Figure \ref{fig_initialpoints} {\bf C}.

\begin{table}[htbp]
    \centering
    \begin{tabular}{c|c c c c}\hline
        $\mathcal{I}$ & $|\overline{\mathcal{E}}|$ & $\mathcal{T}$ & $\overline{\mathcal{L}}$ & $\tilde{\mathcal{T}}$\\\hline
        $(1, 0)^\dagger$ & $2.2520\times10^{-4}$ & $1.4098\times10^{4}$ & $1.0562\times10^3$ & $1.0000$\\
        $(5, 0)$ & $3.6666\times10^{-4}$ & $2.8535\times10^4$ & $7.1597\times10^3$ & $2.4050$\\
        $(0, 5)$ & $4.0497\times10^4$ & $2.9442\times10^4$ & $2.5417\times10^3$ & $3.3887$\\\hline
    \end{tabular}
    \caption{Parameters estimation of initial position $\mathcal{I}$. $^\dagger$The benchmark setup used for parameter tuning.}
    \label{tbale_initialpoints}
\end{table}

The PIDOC estimates corresponding to Figure \ref{fig_initialpoints} are shown in Table \ref{tbale_initialpoints}. It can be deduced from Table \ref{tbale_initialpoints} that the for $\mathcal{I} = (0, 5)$, PIDOC has the highest absolute mean error, with $\mathcal{I} = (1, 0)$ has the lowest. The difference of the $|\overline{\mathcal{E}}|$ for points $(5, 0)$ and $(0, 5)$ may seem conflict with visualization in Figure \ref{fig_initialpoints} {\bf A} and {\bf B}. However, the PIDOC control for $\mathcal{I} = (5,0)$ (pink line in Figure \ref{fig_initialpoints} {\bf A}) has a strong trajectory variation at the initial stage with good convergence to the desired trajectory later, which is also represented in the black dashed line in Figure \ref{fig_initialpoints} {\bf C}, yet for $\mathcal{I} = (0, 5)$ (green line in Figure \ref{fig_initialpoints} {\bf A}) the trajectory fluctuation is evidently more obvious during the whole PIDOC controls, accounts for the higher $|\overline{\mathcal{E}}|$ value in Table \ref{tbale_initialpoints}. For $\mathcal{T}$ and $\tilde{\mathcal{T}}$ such a trend also stands. For $\overline{\mathcal{L}}$, $\mathcal{I} = (5, 0)$ displays an evidently higher value, showing good agreement with Figure \ref{fig_initialpoints} {\bf B}.

\subsubsection{System nonlinearity\label{sec_nonlinearity}}

\begin{figure}[htbp]
    \centering
    \includegraphics[scale=1.1]{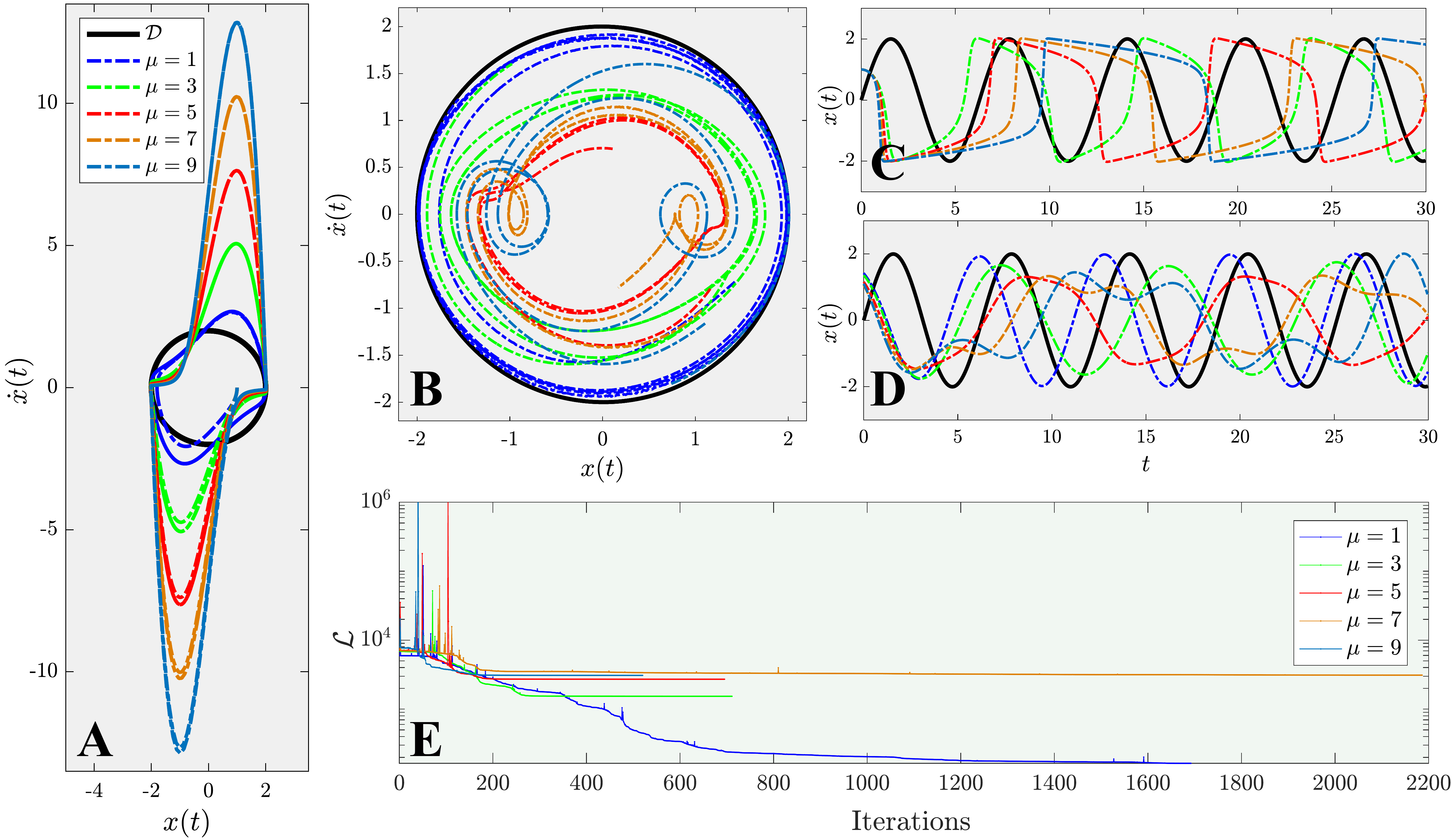}
    \caption{System analysis on PIDOC considering the nonlinearity of the van der Pol systems with different $\mu$. {\bf A}. the inherent dynamics of different van der Pol systems of different $\mu$ marked in dashed lines with different colors with the desired control trajectory $\mathcal{D}$ marked in black. {\bf B}. the phase portrait of the desired trajectory corresponding to PIDOC applied to different systems of different nonlinearities. {C}. the position plot with time $t$ of the system's inherent dynamics. {\bf D}. the position plot with time $t$ of the PIDOC controlled dynamics. {\bf E}. the loss function - iterations plot of PIDOC applied to van der Pol systems of different nonlinearities. Note that the colors used for different PIDOC controls of different nonlinearities all correspond to subfigure {\bf A}.}
    \label{fig_nonlinearity}
\end{figure}

Nonlinearity is the key essence for control, especially for chaotic systems. For van der Pol oscillating circuits, systematic nonlinearities are represented with different $\mu$ values, varying from $\mu = 1,3,5,7,9$, for testing PIDOC controls. Figure \ref{fig_nonlinearity} {\bf A} visualized the phase portrait of such intrinsic nonlinear dynamics. After imposing the PIDOC controls, Figure \ref{fig_nonlinearity} {\bf B} shows the output control with different $\mu$s. Figure \ref{fig_nonlinearity} {\bf C} corresponds to Figure \ref{fig_nonlinearity} {\bf A}, showing how $x(t)$ evolution with time of the van der Pol inherent dynamics; with Figure \ref{fig_nonlinearity} {\bf D} corresponds to Figure \ref{fig_nonlinearity} {\bf D} showing the time evolution of $x(t)$ for PIDOC controls. Figure \ref{fig_nonlinearity} {\bf E} shows the loss functions - iterations diagrams of different PIDOC control for systems of different nonlinearities. 

\begin{figure}[htbp]
    \centering
    \includegraphics{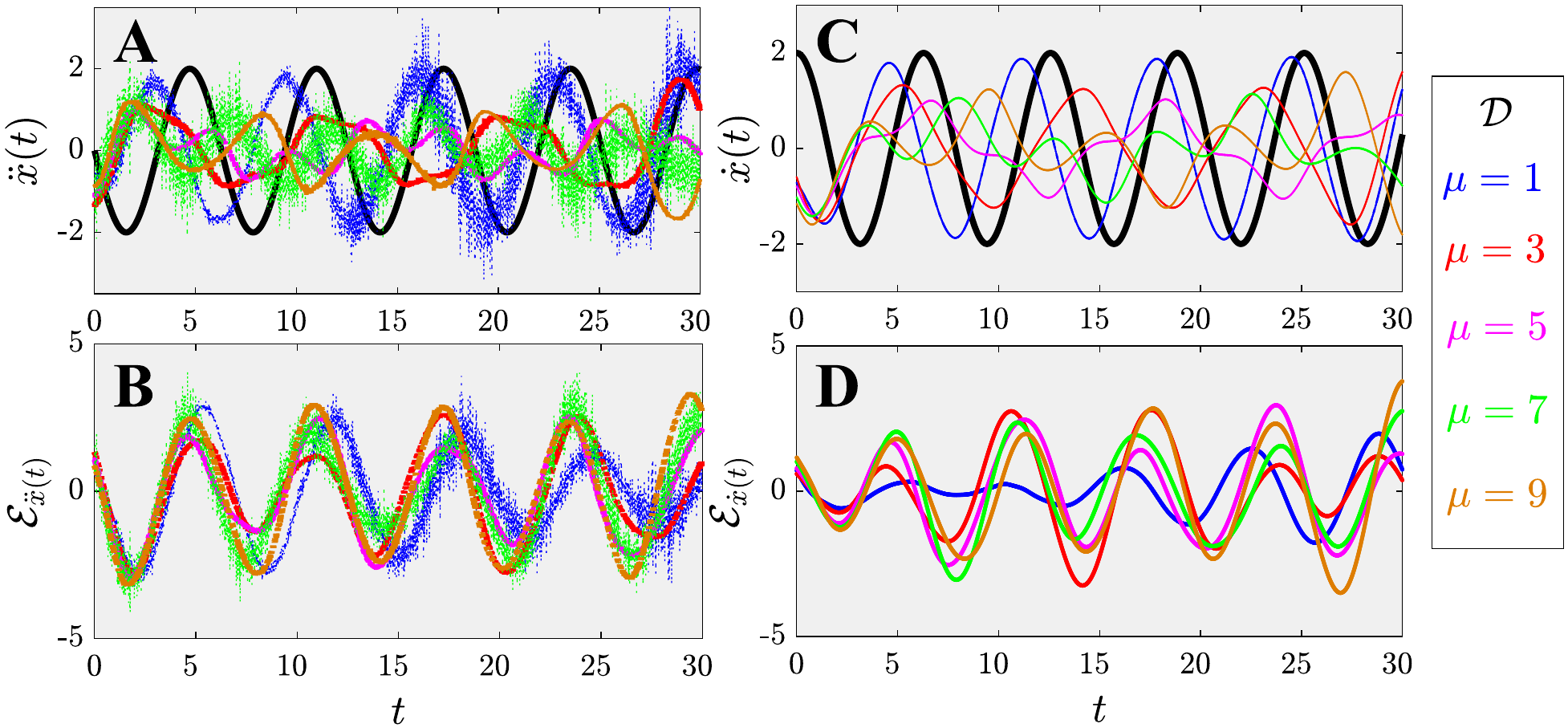}
    \caption{The plot of velocity $\dot{x}(t)$ and acceleration $\ddot{x}(t)$ to their corresponding errors regarding time $t$ considering cases of different nonlinearities. Note that the colors for cases in van der Pol systems of different nonlinearities are shown in the legend on the right side, where $\mathcal{D}$ stands for the desired control trajectory. {\bf A}. the acceleration $\ddot{x}(t)$ - $t$ diagram. {\bf B}. the errors of the acceleration. {\bf C}. the acceleration $\dot{x}(t)$ - $t$ diagram. {\bf D}. the errors of the velocity. }
    \label{fig_nonlinearity_error}
\end{figure}

Figure \ref{fig_nonlinearity} {\bf E} shows an evident lower losses for system with low nonlinearity of $\mu = 1$, along with more iterations with $\mu = 1 \ \&\ 7$. Comparing Figure \ref{fig_nonlinearity_error} {\bf A} and {\bf C} we observe how high nonlinearities in phase portrait displayed on $x(t)$: a lower frequency with a specifc band structure of a "sharpe band shape" indicated in the waves in \ref{fig_nonlinearity_error} {\bf C} (specifically for the blue and orange dashed lines). From Figure \ref{fig_nonlinearity_error} {\bf B} we see a two "inner circles" attached on the two sides for high nonlinearities as can be observed in blue and orange dashed lines. Moving to Figure \ref{fig_nonlinearity_error} {\bf D} a specific "double wave" shape is observed, a smaller wave on a bigger wave as also be observed in blue and orange dashed lines. Comparing Figure \ref{fig_nonlinearity_error} {\bf B} and {\bf D} one can conclude that the "double wave" structure observed contributes to the smaller circles in the phase portrait. Comparing Figure \ref{fig_nonlinearity_error} {\bf D} and {\bf C} one deduce that the reason of the "small wave" generation is the sharp band structure observed in Figure \ref{fig_nonlinearity_error} {\bf C}: the high nonlinearitity generates the sharp wave structure, make imposing the control signal of sinusoidal function significantly difficult.

\begin{table}[htbp]
    \centering
    \begin{tabular}{c|c c c c}\hline
        $\mu$ & $|\overline{\mathcal{E}}|$ & $\mathcal{T}$ & $\overline{\mathcal{L}}$ & $\tilde{\mathcal{T}}$\\\hline
        1$^\dagger$ & $2.2520\times10^{-4}$ & $1.4098\times10^{4}$ & $0.0106\times10^5$ & $1.0000$\\
        $3$ & $2.4732\times10^{-4}$ & $2.3042\times10^{4}$ & $0.0266\times10^5$ & $0.3889$\\
        $5$ & $2.1821\times10^{-4}$ & $2.4628\times10^{4}$ & $0.0579\times10^5$ & $0.4253$\\
        $7$ & $2.4079\times10^{-4}$ & $2.7107\times10^{4}$ & $0.0353\times10^5$ & $ 0.1488$\\
        $9$ & $2.9083\times10^{-4}$ & $3.2397\times10^{4}$ & $2.6962\times10^5$ & $0.7477$\\\hline
    \end{tabular}
    \caption{Parameters estimation of the nonlinearity on $\mu$. $^\dagger$The benchmark setup used for parameter tuning.}
    \label{table_nonlinearity}
\end{table}

Corresponding to Figure \ref{fig_nonlinearity}, Table \ref{table_nonlinearity} shows the PIDOC estimates for different nonlinearities. It is observed that both $\mu=1,3,...,9$, the $|\overline{\mathcal{E}}|$ values are in the same range, with a slight difference in numerical values, which is not expected intuitively based on Figure \ref{fig_nonlinearity}. The $\mathcal{T}$ values increase with nonlinearities. The mean losses $\overline{\mathcal{L}}$ variate but shows an increasing trend with higher nonlinearity, with a loss explosion observed for $\mu = 9$ as explained for Table \ref{table_lagrange}. The mean normalized time $\tilde{\mathcal{T}}$ shows that the computational burden reduces per iteration with higher nonlinearities.

To further explore the similar values of $|\overline{\mathcal{E}}|$, we create Figure \ref{fig_nonlinearity_error}. Figure \ref{fig_nonlinearity_error} {\bf A} and {\bf C} shows the acceleration and velocities of PIDOC controls marked in dashed lines in different colors in the right legend compared with the desired trajectory marked in black solid lines. Figure \ref{fig_nonlinearity_error} {\bf B} and {\bf D} shows the difference (or errors) for velocities and accelerations calculated by \begin{equation}
    \begin{aligned}
    \mathcal{E}_{\ddot{x}(t)} =\ddot{x}_{\mathcal{D}}(t) - \ddot{x}_{pred}(t),\quad \mathcal{E}_{\dot{x}(t)} =\dot{x}_{\mathcal{D}}(t) - \dot{x}_{pred}(t) 
    \end{aligned}
\end{equation}

It is reported that the acceleration errors $\mathcal{E}_{\ddot{x}(t)}$ of different systems with different nonlinearities exhibits the same frequency with the control signal, comparing Figure \ref{fig_nonlinearity_error} {\bf A} and {\bf B}. The errors of velocities $\mathcal{E}_{\dot{x}(t)}$ does not show such evident trends yet all seemingly fluctuating in the same frequency as in Figure \ref{fig_nonlinearity_error} {\bf D}. We can therefore conclude a significant characteristic of PIDOC controls: due to the imposition of $\mathcal{I}$, there will be a phase lag for PIDOC controls, which is reported in Figure \ref{systemanalysis}. Such phase lags generate a so-called error, or difference, between the control signal and PIDOC control. For van der Pol systems of different nonlinearities, the errors exhibit the same frequency, with similar wave range values. The similarities of the range values and frequency combined explain why PIDOC controls exhibit similar errors as reported in Table \ref{table_nonlinearity}. To note, such a phase lag successfully implements the controls for relative low nonlinearities as in Figures \ref{systemanalysis}, yet still exhibits imperfect controls for high nonlinearities.

\subsection{Hyperparameters and control\label{sec_hyperparam}}

\subsubsection{Deep neural networks\label{sec_deepnn}}

To test how the NN structures variate the control process, two cases are set up for investigation: (1) the benchmark problem with reduced neurons and layers, of six different sets: NN structures of $1\times30$, $3\times30$, $6\times30$, $1\times10$, $3\times10$, $6\times10$; and (2) the increased neurons and layers for controlling van der Pol system with high nonlinearity of $\mu = 5$, of five different sets: NN layers of 6, 9, 15, 30, 50, with 30 neurons per layers. The aim of case (1) is to investigate whether reduced neurons and layers in NN, as one expects reduced capabilities of approximating nonlinear data, can still implement controls of high quality, as observed in Figure \ref{systemanalysis}. Particularly, it has been reported by Pinkus \cite{singlelayer} that a single hidden layer NN can approximate nonlinear mappings, followed by a physics-informed practice by Lu {\em et al.} \cite{deepxde}. We therefore specifically test PIDOC with a single hidden layer for testing its ability for controlling the van der Pol system. The aim of the case (2) is to investigate whether increased approximation capacity can tackle the control of highly nonlinear systems as we made effort in Figure \ref{fig_nonlinearity}.

\begin{figure}[htbp]
    \centering
    \includegraphics[scale=1.1]{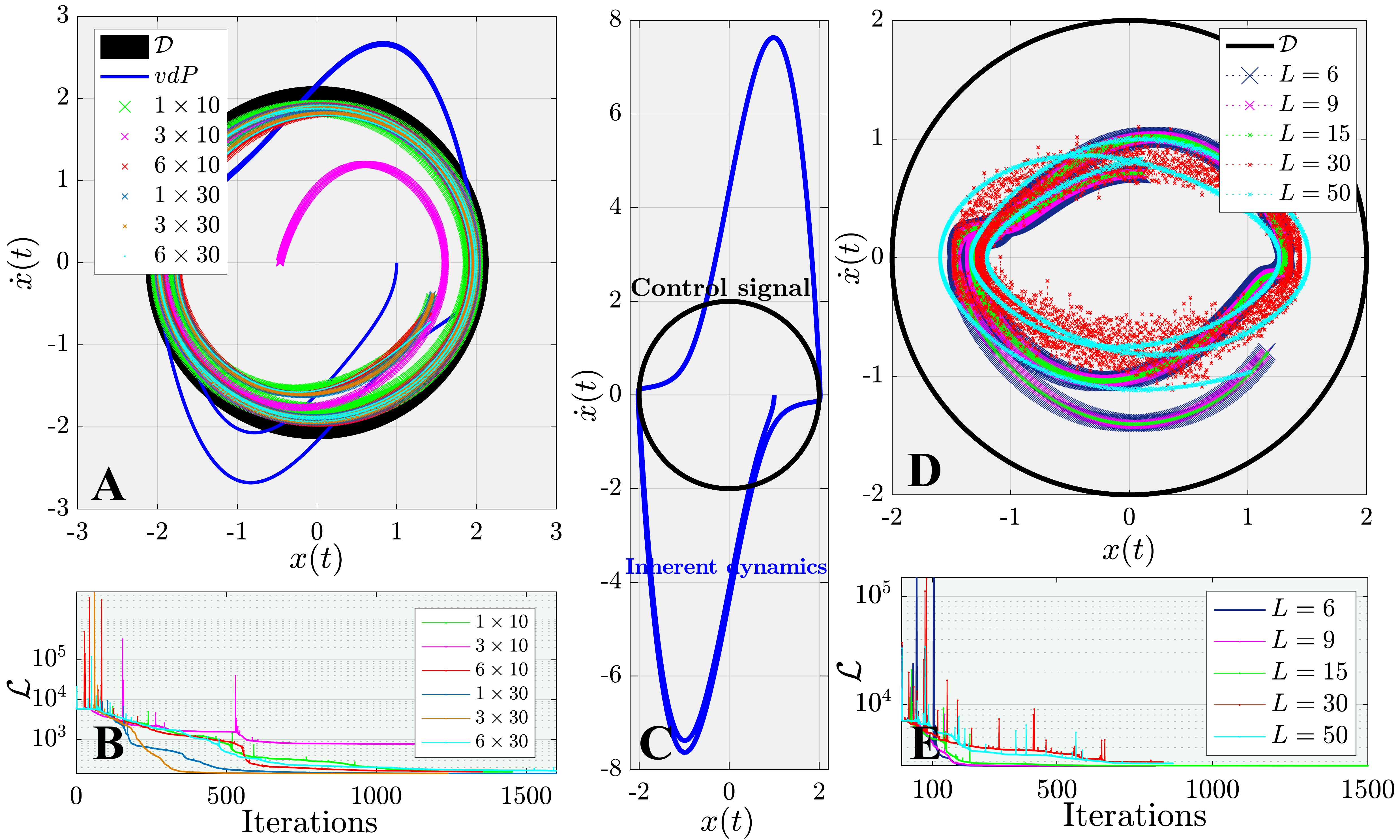}
    \caption{System behavior analysis of how the NN structure tune the control process of PIDOC for $\mu = 1$ and $\mu = 5$. {\bf A}. the phase portrait of PIDOC controls considering different NN structures; the solid black line $\mathcal{D}$ denoted the desired trajectory, the blue solid line $vdP$ denotes the van der Pol inherent dynamics. Different colored cross dots as marked in the legend denotes different NN structures. Note that subfigure {\bf A} and {\bf B} share the same color representation of the NN structure. {\bf B}. the loss function - iterations diagram for different NN structures as applied to the benchmark problem of $\mu = 1$. {\bf C}. the inherent van der Pol dynamics (marked as a blue solid line) when $\mu = 5$. {\bf D}. the phase portrait of PIDOC applied to highly nonlinear van der Pol system of $\mu = 5$ corresponds to subfigure {\bf C}. Note that the black solid line $\mathcal{D}$ is the desired trajectory. Different NN structures are denoted by cross dots in different colors as indicated in the legend. {\bf E}. the loss function - iterations diagram corresponds to subfigure {\bf D}.}
    \label{fig_nnstructure_control}
\end{figure}

Figure \ref{fig_nnstructure_control} {\bf A} and {\bf B} shows for reduced NN layers and neurons all exhibits good control implementations, with a slightly trajectory variation at the beginning stage for NN structure $3 \times 10$ as indicated in the pink line with higher losses. Figure \ref{fig_nnstructure_control} {\bf C} plot the intrinsic dynamics of van der Pol system with $\mu =5$, as comparing with the controlled phase in \ref{fig_nnstructure_control} {\bf D}: with increasing layers the controlled phase shows reduced nonlinearities, especially for layers $L= 6$ and 50 for comparing the light and pure blue dashed lines - the vortex-liked shape on the two sides of the phase when $x \approx 1.2$ evidently reduced with increasing layers. For different layers the losses show a similar trends as reported in Figure \ref{fig_nnstructure_control} {\bf E}. Notably, Figure \ref{fig_nnstructure_control} {\bf A} also indicate that single hidden layer NN shows good approxibilities, with better phase control than NN structure of $3 \times 10$.

\begin{table}[htbp]
    \centering
    \begin{tabular}{c c|c c c c}\hline
        Layers & Neurons & $|\overline{\mathcal{E}}|$ & $\mathcal{T}$ & $\overline{\mathcal{L}}$ & $\tilde{\mathcal{T}}$\\\hline
        1 & 30 & $2.2472\times10{-4}$ & $1.2558\times10^4$ & $691.4871$ & $0.9229$\\ 
        3 & 30 & $2.2474\times10{-4}$ & $1.2853\times10^4$ & $3.8015\times10^5$ & $1.2450$\\
        6$^\dagger$ & 30$^\dagger$ & $2.2520\times10{-4}$ & $1.1098\times10^4$ & $1.0562\times10^3$ & $1.0000$\\
        1 & 10 & $2.2441\times10{-4}$ & $1.0823\times10^4$ & $1.1010\times10^3$ & $0.7572$\\
        3 & 10 & $2.1779\times10{-4}$ & $1.1285\times10^4$ & $1.9470\times10^3$ & $1.0958$\\
        6 & 10 & $2.2439\times10{-4}$ & $1.1444\times10^4$ & $6.6087\times10^3$ & $1.0144$\\\hline
    \end{tabular}
    \caption{Parameters estimation of the NN structure considering layers and neurons to the benchmark setup. $^\dagger$The benchmark setup used for parameter tuning where $\mu=1$.}
    \label{table_nnstructure_control}
\end{table}

Numerical investigation of Figure \ref{fig_nnstructure_control} {\bf A} and {\bf B} represented by PIDOC estimates for $\mu =1$ are shown in Table \ref{table_nnstructure_control}. The $|\overline{\mathcal{E}}|$ and $\mathcal{T}$ are in approximately the same range for different NN structures. The losses evidently higher for NN of $3\times30$ and $6 \times10$. The higher losses for NN of $3\times10$ can be captured in Figure \ref{fig_nnstructure_control} {\bf B}; yet for bigger $\mathcal{L}$ for NN of $3\times30$ and $6 \times10$, we can account it for the high loss fluctutations at the beginning stage, as also reported Table \ref{table_lagrange}. For $\tilde{\mathcal{T}}$ we reports NN structure $3\times30$ took higher training time per iterations, and for $3\times10$ and $6\times10$ the $\tilde{\mathcal{T}}$ values are slightly higher.

\begin{table}[htbp]
    \centering
    \begin{tabular}{c c|c c c c}\hline
        Layers & Neurons & $|\overline{\mathcal{E}}|$ & $\mathcal{T}$ & $\overline{\mathcal{L}}$ & $\tilde{\mathcal{T}}$\\\hline
        6$^\dagger$ & 30$^\dagger$ & $2.3083\times10^{-4}$ & $1.9409\times10^4$ & $3.1732\times10^3$ & $1.0000$\\
        9 & 30 & $2.3091\times10^{-4}$ & $1.9591\times10^4$ & $8.6195\times10^3$ & $6.9859$\\
        15 & 30 & $2.3055\times10^{-4}$ & $2.0028\times10^4$ & $3.4388\times10^3$ & $5.4510$\\
        30 & 30 & $2.2702\times10^{-4}$ & $2.0759\times10^4$ & $8.2714 \times10^3$ & $3.4597$\\
        50 & 30 & $2.2431\times10^{-4}$ & $2.1228\times10^4$ & $3.9537\times10^3$ & $33.7096$\\\hline
    \end{tabular}
    \caption{Parameters estimation of the NN structure regarding layers and neurons, estimating a highly nonlinear van der Pol system with $\mu =5$. $^\dagger$The benchmark NN structure used for parameter tuning, adopting the van der Pol system of high nonlinearities.}
    \label{table_nnstructure_control_mu5}
\end{table}

The PIDOC estimates of highly nonlinear van der Pol system of $\mu = 5$ are shown in Table \ref{table_nnstructure_control_mu5}. The values $|\overline{\mathcal{E}}|$ are basically in the same range. A generally higher training time $\mathcal{T}$ and normalized training time per iterations $\tilde{\mathcal{T}}$ are reported for increasing layers. The increase on $\tilde{\mathcal{T}}$ indicates the optimization described in Equation \ref{opt} stops earlier with more layers, resulting in fewer iterations. 

\subsubsection{Lagrangian multiplier\label{sec_lagran}}

\begin{figure}[htbp]
    \centering
    \includegraphics{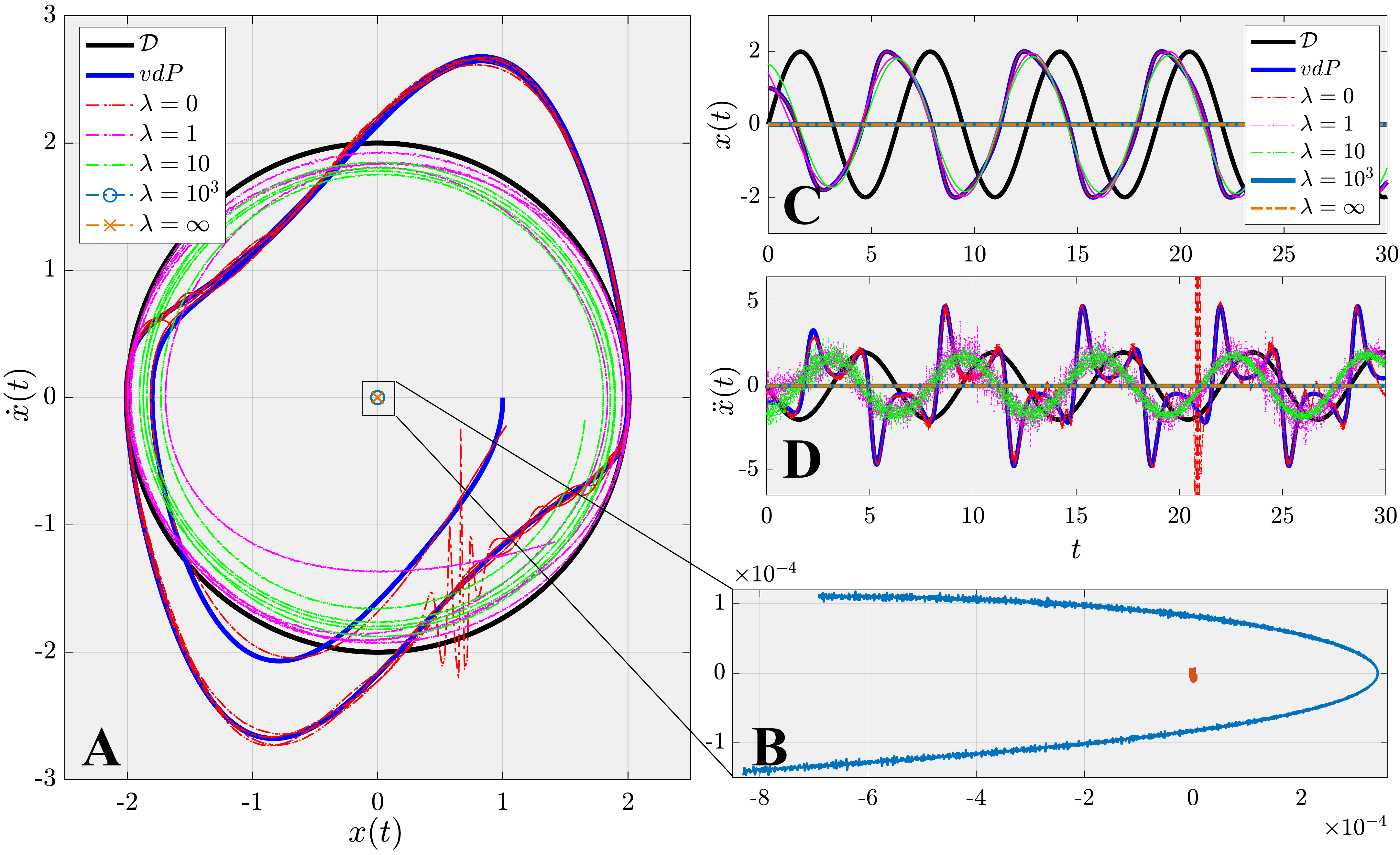}
    \caption{System behavior analysis of the effects of the Lagrangian multiplier on the control signal implementation for PIDOC. Note that the black solid line $\mathcal{D}$ is the desired trajectory, the blue solid line $vdP$ denotes the van der Pol inherent dynamics, and the controlled route of different Lagrangian multiplier values are denoted in different colored dashed lines. Note that all the colors in the subfigures are marked the same as represented in the legend in subfigure {\bf A}. {\bf A}. the phase portrait of the PIDOC controls. {\bf B}. zoomed view for rescale for Lagrangian multiplier $\lambda = 10^3$ and $\lambda = \infty$. {\bf C}. the position $x(t)$ regarding time $t$ for desired trajectory $\mathcal{D}$, van der Pol inherent dynamics and different PIDOC controls. {\bf D}. the acceleration $\ddot{x}(t)$ regarding time $t$ for desired trajectory $\mathcal{D}$, van der Pol inherent dynamics and different PIDOC controls. }
    \label{fig_lagrangianmultiplier}
\end{figure}

It is intuitive to think that by enlarging the control signal we might expect a better control implementation. Curious about the effects of control signals, we applied different Lagrangian multiplier $\lambda = 0, 1, 10, 10^3, \infty$, for testing the systematic control accuracy. For $\lambda = 0$, there are simply no control signals encoded in the loss, where PIDOC is reduced to a normal NN with only the initial position encoded as a soft constraint. The problem thence turned into a standard NN learning and fitting problem. For $\lambda = \infty$, we simply simply eliminate Equations (\ref{initial_loss}) and (\ref{desire_loss}), as the control signals turn into infinity. The phase portrait, the time evolution of position $x(t)$ and $\ddot{x}(t)$ of different $\lambda$s are shown in Figure \ref{fig_lagrangianmultiplier}.

Figure \ref{fig_lagrangianmultiplier} {\bf A} shows the phase portrait of PIDOC controls with different $\lambda$s, indicates the pure NN learning of van der Pol dynamics displays display a fluctuation at the area around $[x(t), \dot{x}(t)] \approx [0.5, -1.5]$ for the red dashed line. The pink and green lines indicate for $\lambda = 1$ \& 10 PIDOC implements generally good controls as both converge to the circular trajectories. However, as shown in the zoomed view in Figure \ref{fig_lagrangianmultiplier} {\bf B} one can discern with very high $\lambda$ the PIDOC is difunctionalized as the controlled trajectory shrink to a very low value ($\approx 10^{-4}$). Figure \ref{fig_lagrangianmultiplier} {\bf C} and {\bf D} are the positions $x(t)$ and accelerations $\ddot{x}(t)$ for different weights of control signals ($\lambda$). Both the subfigures indicate when $\lambda$ approach a high value ($10^3$ \& $\infty$) the positions and accelerations shrink to a very low value as can be observed from the blue and orange lines, corresponds to Figure \ref{fig_lagrangianmultiplier} {\bf A}. Notably, it has also been observed that a robust acceleration fluctuation occurs at $t\approx 21$ in the red dashed line, corresponds to the phase fluctuations in Figure \ref{fig_lagrangianmultiplier} {\bf A}, as the errors of NN approximations. Such errors can also be attributed to the stochastic nature of NN as we previously explained for Figure \ref{systemanalysis}.

\begin{table}[htbp]
    \centering
    \begin{tabular}{c|c c c c}\hline
        $\lambda$ & $|\overline{\mathcal{E}}|$ & $\mathcal{T}$ & $\overline{\mathcal{L}}$ & $\tilde{\mathcal{T}}$\\\hline
        0 & $1.0181\times10^{-4}$ & $7.0409\times10^{3}$& $415.1019$ & $0.6867$\\
        1$^\dagger$ & $1.0221 \times10^{-4}$ & $6.2442\times10^{3}$ & $5835.0830$ & $1.0000$\\
        $10$ & $0.8626\times10^{-4}$ & $6.4437\times10^{3}$ &  $2419.3699$ & $0.9709$\\
        $10^3$ & $2.1127\times10^{-4}$ & $6.5846\times10^{3}$ &  $45516.0465$ & $47.2085$\\
        $\infty$ & $2.1127\times10^{-4}$ & $6.7030\times10^{3}$ & $3.7753$ & $52.9858$\\\hline
    \end{tabular}
    \caption{Parameters estimation of the Lagrangian multiplier $\lambda$ or enlarging or eliminating the effects of the control signal in the physics-informed loss. $^\dagger$The benchmark setup used for parameter tuning.}
    \label{table_lagrangianmultiplier}
\end{table}

Table \ref{table_lagrangianmultiplier} numerically unveils how the weights of control signals affect PIDOC estimates. From the values of $|\overline{\mathcal{E}}|$ one can deduce for $\lambda = 10^3$ and $\infty$ there are an evidently higher errors. The training time $\mathcal{T}$ basically holds the same for both cases. However, it should be noted that there is an increasing $\overline{\mathcal{L}}$ as $\lambda$ increases, yet when the control signal is eliminated, $\overline{\mathcal{L}}$ reduced to a very low value, as the problem turned into pure NN learning. The $\tilde{\mathcal{T}}$ values increases significantly for $\lambda = 10^3$ and $\infty$, which connected with Figure \ref{fig_lagrangianmultiplier} {\bf A} and {\bf B} indicating increasing computational burden per iterations and low quality control caused by the high weights of control signals in PIDOC. From such results we can further propose an explanation for implementations of physics-informed controls: the high weights of control signals lead to the deprivation of information of the training data. Such deprivation may "confuse" the learning of NN as it is mainly designed for stochastic data-based learning and shows robust capabilities given a humongous dataset given no external constraint \cite{deeplearning, george}. Hence, as the core of deep learning, even the goal emphasizes control, the given data is always of key essence. We hence conclude that even the given training data of the van der Pol system is nonlinear it still contributes largely to the successful implementation of PIDOC controls. 

\section{Concluding Remarks and Future Works\label{sec_conclusion}}

In this paper we tackle a century-old yet widely applied problem, controlling a nonlinear van der Pol dynamical system, with a novel approach using Physics-Informed Neural Networks. Instead of adopting the traditional paradigm of learning and predicting using PINN, we use PINN for controlling nonlinear systems. A new PINN-based framework \textsc{Physics-Informed Deep Operator Control}, shortened as PIDOC, is presented, consisting of a deep neural network and the physics-informed control, including the desired control trajectories and initial positions. PIDOC is fed with systematic nonlinear data to control van der Pol circuits to output the controlled signals. To investigate the behavior and properties of PIDOC, we first applied PIDOC for benchmark control problems for systematic analysis, then designed three sets of numerical experiments for testing the effects of amplitudes of desired trajectories $\Lambda$, different initial points $\mathcal{I}$, and system nonlinearities as represented by $\mu$. We then tune the hyperparameters to change the neurons and layers of the NN to study two problems: (1) does a NN with smaller volume still shows the same capability of controlling chaos applied to the benchmark problem; (2) can increase NN volume demonstrate better capabilities on controlling van der Pol systems with high nonlinearities. We also intend to verify the capability of single hidden layer NN to approximate nonlinear systems for part of the control. We also change Lagrangian multiplier $\lambda$ as a weight factor to check how desired trajectories as control signals guide PIDOC in the control process. 

Results indicate PIDOC controls exhibit higher stochasticity for higher-order terms, as can be attributed to the stochastic nature of deep learning, with a successful implementation of the desired trajectory on the benchmark problem. PIDOC also demonstrates capacity on increased trajectory amplitudes with lower absolute mean errors. For systems with different initial points, numerical experiments show for points further away PIDOC can still successfully implement controls as higher fluctuations at the initial stage. However, as we increase the system nonlinearities, the PIDOC output controls are not as ideal as the benchmark problems, as two vortex-liked structures occur on the phase portrait, with an evidently higher loss for systems with high nonlinearities. A decreased NN in PIDOC volume also shows good control implementations with the van der Pol system of $\mu = 1$, while increasing the layers did not make systems of high nonlinearity with $\mu = 5$ follow the desired trajectory as well as the benchmark problem. It should be noted that increasing layers do generate improvement on the output controlled signals as the vortex-liked structures in phase portrait vanished, making the system more predictive. Increasing the weights of the control signals in PIDOC does not improve the control qualities based on the output. A further conclusion is made, even the systematic data is nonlinear and chaotic, it still contributes much to the PIDOC controls as PIDOC is intrinsically a deep learning-based control method.

Considering the successful implementation of PIDOC to control different van der Pol systems, further investigation on using PIDOC to impose control to other systems such as the Lorentz system would have brought in more insights into PIDOC. Also, comparison on the control properties based on PIDOC and deterministic controls, i.e., Cooper {\em et al.} \cite{cooper} could also be potential researches. Improvement on PIDOC or further developed models tackling systems with high nonlinearities is of significance for future research.

\section*{Data Availability}

All the code and data will be made publicly available upon acceptance of the manuscript through \url{https://github.com/hanfengzhai/PIDOC}. For the original PINNs code please refer to \url{https://github.com/maziarraissi/PINNs}. The training of the deep learning algorithms and simulation of the van der Pol system is conducted on \href{https://colab.research.google.com/}{Google Colab} \cite{colab}.


\end{document}